\documentclass{LMCS}
\usepackage{macros,enumerate,hyperref}
\let\subst\subone
\showcomments

\title{\lrbac: Programming with Role-Based Access Control}
\author[R.~Jagadeesan]{Radha Jagadeesan\rsuper a}
\address{{\lsuper{a,c,d}}CTI, DePaul University}
\email{\{rjagadeesan,cpitcher,jriely\}@cti.depaul.edu}
\thanks{{\lsuper{a,c}}Radha Jagadeesan and Corin Pitcher were 
supported in part by NSF CyberTrust 0430175.}
\author[A.~Jeffrey]{Alan Jeffrey\rsuper b}
\address{{\lsuper b}Bell Labs}
\email{ajeffrey@bell-labs.com}
\thanks{}
\author[C.~Pitcher]{Corin Pitcher\rsuper c}
\address{\vskip-6 pt}
\thanks{}
\author[J.~Riely]{James Riely\rsuper d}
\address{\vskip-6 pt}
\thanks{{\lsuper d}James Riely was supported in part by NSF CAREER 0347542.}
\def\doi{4 (1:2) 2008}
\lmcsheading%
{\doi}
{1--24}
{}
{}
{Nov.~21, 2006}
{Jan.~\phantom{0}9, 2008}
{}   

\begin{document}

%
%
%
\begin{abstract}
  We study mechanisms that permit program components to express role
  constraints on clients, focusing on programmatic security
  mechanisms, which permit access controls to be expressed, \emph{in
    situ}, as part of the code realizing basic functionality.  In this
  setting, two questions immediately arise.  (1) The user of a
  component faces the issue of safety: is a particular role sufficient
  to use the component? (2) The component designer faces the dual
  issue of protection: is a particular role demanded in all execution
  paths of the component? We provide a formal calculus and static
  analysis to answer both questions.
\end{abstract}
\maketitle

%
\section{Introduction}
\label{sec:begin}

This paper addresses programmatic security mechanisms as realized in
systems such as Java Authentication and Authorization Service (\JAAS)
and \DOTNET.  These systems enable two forms of access control
mechanisms\footnote{In this paper, we discuss only authorization
  mechanisms, ignoring the authentication mechanisms that are also
  part of these infrastructures.}.  First, they permit
\emph{declarative} access control to describe security
specifications that are orthogonal and separate from descriptions of
functionality, e.g., in an interface $I$, a declarative access
control mechanism could require the caller to possess a minimum set
of rights.  While conceptually elegant, such specifications do not
directly permit the enforcement of access control that is sensitive
to the control and dataflow of the code implementing the
functionality --- consider for example history sensitive security
policies that require runtime monitoring of relevant events.
Consequently, \JAAS\ and \DOTNET\ also include \emph{programmatic}
mechanisms that permit access control code to be intertwined with
functionality code, e.g., in the code of a component implementing
interface $I$. On the one hand, such programmatic mechanisms permit
the direct expression of access control policies.  However, the
programmatic approach leads to the commingling of the conceptually
separate concerns of security and functionality.

There is extensive literature on policy languages to specify and
implement policies
(e.g.,~\cite{DBLP:journals/ijisec/LigattiBW05,1035433,383894,950194,507715,implementing-rbac:hoffman}
to name but a few).  This research studies security policies as
separate and orthogonal additions to component code, and is thus
focused on declarative security in the parlance of \JAAS/\DOTNET.

In contrast, we study programmatic security mechanisms.  Our
motivation is to {\em extract} the security guarantees provided by
access control code which has been written inline with component
code. We address this issue from two viewpoints:
\begin{enumerate}[$\bullet$]
\item The user of a component faces the issue of safety: is a
  particular set of rights sufficient to use the component? (ie. with
  that set of rights, there is no possible execution path that would
  fail a security check.  Furthermore, any
  greater set of rights will also be allowed to use the component)
\item The component designer faces the dual issue of protection: is a
  particular set of rights demanded in all execution paths of the
  component? (ie. every execution path requires that set of rights.  Furthermore,
  any lesser set of rights will not be allowed to use
  the component)
\end{enumerate}
The main contribution of this paper is separate static analyses to
calculate approximations to these two questions. An approximate answer
to the first question is a set of rights, perhaps bigger than
necessary, that is \emph{sufficient} to use the component. On the
other hand, an approximate answer to the second question, is a set of
rights, perhaps smaller than what is actually enforced, that is
\emph{necessary} to use the component.

%
\subsection{An overview of our technical contributions}

There is extensive literature on Role-Based Access-Control (\RBAC)
models including \NIST\ standards for \RBAC~\cite{Orig-RBAC,NIST-RBAC};
see~\cite{rbac-book} for a textbook survey.  The main motivation for
\RBAC, in software architectures (e.g.,~\cite{Web-RBAC,DAC-RBAC}) and
frameworks such as \JAAS/\DOTNET, is that it enables the enforcement
of security policies at a granularity demanded by the application. In
these examples, \RBAC\ allows permissions to be de-coupled from users:
Roles are the unit of administration for users and permissions are
assigned to roles.  Roles are often arranged in a hierarchy for
succinct representation of the mapping of permissions.  Component
programmers design code in terms of a static collection of roles.
When the application is deployed, administrators map the roles defined
in the application to users in the particular domain.

In this paper, we study a lambda calculus enriched with primitives for
access control, dubbed \lrbac.  The underlying lambda calculus serves
as an abstraction of the ambient programming framework in a real
system.  We draw inspiration from the programming idioms in \JAAS\ and
\DOTNET, to determine the expressiveness required for the access
control mechanisms.  In a sequence of \DOTNET\ examples\footnote{In
  order to minimize the syntactic barrage on the unsuspecting reader,
  our examples to illustrate the features are drawn solely from the
  \DOTNET\ programming domain.  At the level of our discussion, there
  are no real distinctions between \JAAS\ and \DOTNET\ security
  services.}, closely based on~\cite{Malhotra02}, we give the reader a
flavor of the basic programming idioms.
\begin{example}[~\cite{Malhotra02}]
  \label{ex:check}
  In the \DOTNET\ Framework \CLR, every thread has a \verb|Principal| object that
  carries its role.  This \verb|Principal| object can be viewed as
  representing the user executing the thread. In programming, it often needs to be determined whether
  a specific \verb|Principal| object belongs to a familiar role. The code
  performs checks by making a security call for a \verb|PrincipalPermission|
  object. The \verb|PrincipalPermission| class denotes the role that a
  specific principal needs to match.  At the time of a security check, the
  \CLR\ checks whether the role of the \verb|Principal| object of the caller
  matches the role of the \verb|PrincipalPermission| object being requested.
  If the role values of the two objects do not match, an exception is
  raised. The following code snippet illustrates the issues:
\begin{verbatim}
  PrincipalPermission usrPerm  =
       new PrincipalPermission (null,"Manager");
  usrPerm.Demand()
\end{verbatim}
  If the current thread is associated with a principal that has the the role
  of manager, the \verb|Principal|\-\verb|Permission| objects are created and
  security access is given as required. If the credentials are not valid,
  a security exception is raised. \qed
\end{example}
In this vein, the intuitive operation of \lrbac\ is as follows.
\lrbac\ program execution takes place in the context of a role, say
$r$, which can be viewed concretely as a set of permissions. The set
of roles used in a program is static: we do not allow the dynamic
creation of roles. \lrbac\ supports run-time operations to create
objects (i.e. higher-order functions) that are wrapped with
protecting roles.  The use of such guarded objects is facilitated by
operations that check that the role-context $r$ is at least as
strong as the guarding role: an exception is raised if the check
fails.

The next example illustrates that boolean combinations of roles
are permitted in programs.  In classical \RBAC\ terms, this is
abstracted by a lattice or boolean structure on roles.
\begin{example}[~\cite{Malhotra02}]
  The \verb|Union| method of the \verb|PrincipalPermission| class combines
  multiple \verb|PrincipalPermission| objects. The following code represents
  a security check that succeeds only if the \verb|Principal| object
  represents a user in the \verb|CourseAdmin| or \verb|BudgetManager| roles:
\begin{verbatim}
 PrincipalPermission Perm1 =
     new PrincipalPermission (null,"CourseAdmin");
 PrincipalPermission Perm2 =
     new PrincipalPermission(null,"BudgetManager');

 // Demand at least one of the roles using Union
 perm1.Union (perm2).Demand ()
\end{verbatim}
  Similarly, there is an \verb|Intersect| method to represent a ``join''
  operation in the role lattice.
  \qed
\end{example}
In \lrbac, we assume that roles form a lattice: abstracting the
concrete union/inter\-sec\-tion operations of these examples.  In the
concrete view of a role as a set of permissions, role ordering is
given  by supersets, ie. a role is stronger than another role if it
has more permissions; join of roles corresponds to the union of the
sets of permissions and meet of roles corresponds to the
intersection of the sets of permissions. Some of our results assume
that the lattice is boolean, i.e. the lattice has a negation
operation.  In the concrete view of the motivating examples, the
negation operation is interpreted by set complement with respect to
a maximum collection of permissions

Our study is parametric on the underlying role lattice.

The key operation in such programming is {\em rights modulation}.
From a programming viewpoint, it is convenient, indeed sometimes
required, for an application to operate under the guise of
different users at  different times. Rights modulation of course
comes in two flavors: rights weakening is overall a safe
operation, since the caller chooses to execute with fewer rights.
On the other hand, rights amplification is clearly a more
dangerous operation. In the \DOTNET\ framework, rights modulation is
achieved via a technique called impersonation.
\begin{example}
  \label{ex:impersonate}
  Impersonation of an account is achieved 
  using
  the account's token, as shown in the following code snippet:
\begin{verbatim}
WindowsIdentity stIdentity = new WindowsIdentity (StToken);
 // StToken is the token associated with the Windows acct being impersonated
WindowsImpersonationContext stImp = stIdentity.Impersonate();
 // now operating under the new identity
stImp.Undo();  // revert back
\end{verbatim}
\qed
\end{example}
\lrbac\ has combinators to perform scoped rights weakening and
amplification.

We demonstrate the expressiveness of \lrbac\ by building a range
of useful combinators and a variety of small illustrative
examples. We discuss type systems to perform the two analyses
alluded to earlier: (a) an analysis to detect and remove
unnecessary role-checks in a piece of code for a caller at a
sufficiently high role, and (b) an analysis to determine the
(maximal) role that is guaranteed to be required by a piece of
code. The latter analysis acquires particular value in the
presence of rights modulation.  For both we prove preservation and
progress properties.

%
\subsection{Related work}
Our paper falls into the broad area of research enlarging the
scope of foundational, language-based security methods
(see~\cite{Schneider+:Language-based:Security,Mitchellsurvey,AbadiMS05}
for surveys).

Our work is close in spirit, if not in technical development, to
edit automata~\cite{DBLP:journals/ijisec/LigattiBW05}, which use
aspects to avoid the explicit intermingling of security and
baseline code.

The papers that are most directly relevant to the current paper
are those of Braghin, Gorla and Sassone~\cite{BraghinGS04} and
Compagnoni, Garalda and Gunter ~\cite{Adriana2005}.
~\cite{BraghinGS04} presents the first concurrent calculus with a
notion of \RBAC, whereas~\cite{Adriana2005}'s language enables
privileges depending upon location.

Both these papers start off with a mobile process-based
computational model. Both calculi have primitives to activate and
deactivate roles: these roles are used to prevent undesired
mobility and/or communication, and are similar to the primitives
for role restriction and amplification in this paper.  The ambient
process calculus framework of these papers provides a direct
representation of the ``sessions'' of \RBAC --- in contrast, our
sequential calculus is best thought of as modeling a single
session.

\cite{BraghinGS04,Adriana2005} develop type systems to provide
guarantees about the minimal role required for execution to be
successful --- our first type system occupies the same conceptual
space as this static analysis. However, our second type system
that calculates minimum access controls does not seem to have an
analogue in these papers.

More globally, our paper has been influenced by the desire to
serve loosely as a metalanguage for programming \RBAC\ mechanisms
in examples such as the \JAAS/\DOTNET\ frameworks. Thus, our
treatment internalizes rights amplification by program combinators
and the amplify role constructor in role lattices. In contrast,
the above papers use external --- i.e. not part of the process
language
--- mechanisms (namely, user policies in~\cite{Adriana2005}, and
\RBAC-schemes in~\cite{BraghinGS04}) to enforce control on rights
activation.   We expect that our ideas can be adapted to the
process calculi framework. In future work, we also hope to
integrate the powerful bisimulation principles of these papers.

Our paper deals with access control, so the extensive work on
information flow, e.g., see~\cite{Sabelfeld:Myers:JSAC} for a
survey, is not directly relevant.  However, we note that rights
amplification plays the same role in \lrbac\ that declassification
and delimited
release~\cite{DBLP:conf/ccs/ChongM04,DBLP:conf/isss2/SabelfeldM03,DBLP:conf/csfw/MyersSZ04}
plays in the context of information flow; namely that of
permitting access that would not have been possible otherwise. In
addition, by supporting the internalizing of the ability to
amplify code rights into the role lattice, our system permits
access control code to actively participate in managing rights
amplification.


%
\subsection{Rest of the paper}

We present the language in \autoref{sec:language}, the type system
in \autoref{sec:typing} and illustrate its expressiveness with
examples in \autoref{sec:examples}. We discuss methods for controlling
rights amplification in \autoref{sec:usercode}.  \autoref{sec:proofs} provides
proofs of the theorems from \autoref{sec:typing}.

%
%
%
\newcommand{\seatob}{from\OP<A\OP,B\OP>}%
\newcommand{\secheckb}{test\OP<B\OP>}%
\newcommand{\seatobprime}{from'\OP<A\OP,B\OP>}%
\newcommand{\secheckbprime}{test'\OP<B\OP>}%

%
\section{The Language}
\label{sec:language}

After a discussion of roles, we present an overview of the language
design.  The remaining subsections present the formal syntax,
evaluation semantics, typing system, and some simple examples.

%
\subsection{Roles}
\label{sec:roles}

The language of roles is built up from \emph{role constructors}. The
choice of role constructors is application dependent, but must include
the lattice constructors discussed below.  Each role constructor,
$\arcon$, has an associated arity, $\arity{\arcon}$.  Roles
$\arol\THROUGH\erol$ have the form $\arconto{\arol}$.

We require that roles form a boolean lattice; that is, the set of
constructors must include the nullary constructors $\rleast$ and
$\rmost$, binary constructors $\rmore$ and $\rless$ (written infix),
and unary constructor $\rnegate{}$ (written postfix). $\rleast$ is the
least element of the role lattice. $\rmost$ is the greatest element.
$\rless$ and $\rmore$ are idempotent, commutative, associative, and
mutually distributive meet and join operations respectively.
$\rnegate{}$ is the complement operator.

A role may be thought of as a set of permissions.  Under this
interpretation, $\rleast$ is the empty set, while $\rmost$ is the set
of all permissions.

The syntax of terms uses \emph{role modifiers}, $\arm$, which may be
of the form $\rmup{\arol}{}$ or $\rmdn{\arol}{}$.  We use role modifiers
as functions from roles to roles, with $\arm[\arol]$ defined as follows:
\begin{align*}
  \rmup{\arol}{\brol} &=\arol\rmore\brol
  &
  \rmdn{\arol}{\brol} &=\arol\rless\brol
\end{align*}

In summary, the syntax of roles is as follows.
\smallskip
\begin{displaytab}[\renewcommand{\displayratio}{.40}]{}
  \categoryonedots{\arcon}{Role constructors}
  {{\rleast}\BNFSEP
    {\rmost}\BNFSEP
    {\rmore}\BNFSEP
    {\rless}\BNFSEP
    {\rnegate{}}
  }
  \\[1ex]
  \categoryone{\arol\THROUGH\erol}{\arconto{\arol}}{Roles} 
  \\[1ex]
  \categoryone{\arm}{\rmup{\arol}{} \BNFSEP \rmdn{\arol}{}}{Role modifiers}
\end{displaytab}
Throughout the paper, we assume that all roles (and therefore all
types) are well-formed, in the sense that role constructors have the
correct number of arguments.

The semantics of roles is defined by the relation
``$\okreq{}{\arol}{\brol}$'' stating that $\arol$ and $\brol$ are
provably equivalent.
In addition to any application-specific axioms, we assume
the standard axioms of boolean algebra\nocomment{%
  \begin{math}
    \renewcommand{\arraycolsep}{1pt}
    \begin{array}[t]{rl>{\quad}rll>{\quad}l}
      \arol\rmore(\brol\rmore \crol)&\req (\arol\rmore \brol)\rmore \crol&
      \arol\rless(\brol\rless \crol)&\req (\arol\rless \brol)\rless \crol&
      &\text{Associativity}
      \\
      \arol\rmore \brol&\req\brol\rmore \arol &
      \arol\rless \brol&\req\brol\rless \arol &
      &\text{Commutativity}
      \\
      \arol \rmore (\arol\rless \brol)&\req\arol &
      \arol \rless (\arol\rmore \brol)&\req\arol &
      &\text{Absorption}
      \\
      \arol\rmore (\brol\rless \crol)&\req(\arol\rmore \brol)\rless(\arol\rmore \crol) &
      \arol\rless (\brol\rmore \crol)&\req(\arol\rless \brol)\rmore(\arol\rless \brol) &
      &\text{Distributivity}
      \\
      \arol\rmore\rnegate\arol&\req\rmost &
      \arol\rless\rnegate\arol&\req\rleast &
      &\text{Complement}
    \end{array}
  \end{math}}. %
We say that $\arol$ \emph{dominates} $\brol$ (notation
$\okrsup{}{\arol}{\brol}$) if $\arol\req\arol\rmore\brol$
(equivalently $\brol\req\arol\rless\brol$) is
derivable.
%
Thus we can conclude
$\rmost
\rsup
\arol\rmore\brol
\rsup
\arol
\rsup
\arol\rless\brol
\rsup
\rleast$, for any $\arol$, $\brol$.
%

The role modifier $\rmdn{\arol}{}$ creates a weaker role (closer to
$\rleast$), thus we refer to it as a \emph{restriction}.  Dually, the
modifier $\rmup{\arol}{}$ creates a stronger role (closer to
$\rmost$), and thus we refer to it as an \emph{amplification}.  While
this ordering follows that of the \textsc{nist rbac} standard
\cite{NIST-RBAC}, it is dual to the normal logical reading; it may be
helpful to keep in mind that, viewed as a logic, $\rmost$ is
``false'', $\rleast$ is ``true'', $\rmore$ is ``and'', $\rless$ is
``or'' and $\rsup$ is ``implies.''

%
\subsection{Language overview}
\label{sec:overview}

Our goal is to capture the essence of role-based systems, where roles
are used to regulate the interaction of components of the system.  We
have chosen to base our language on Moggi's monadic metalanguage
because it is simple and well understood, yet rich enough to capture
the key concepts.  By design, the monadic metalanguage is particularly
well suited to studying computational side effects (or simply
\emph{effects}), which are central to our work.  (We expect that our
ideas can be adapted to both process and object calculi.)

The ``components'' in the monadic metalanguage are terms and the
contexts that use them.  To protect terms, we introduce guards
of the form $\pgrd{\arol}{\atrm}$, which can only be discharged by a
context whose role dominates $\arol$.  The notion of \emph{context
  role} is formalized in the definition of evaluation, where
$\okeval{\arol}{\atrm}{\btrm}$ indicates that context role $\arol$ is
sufficient to reduce $\atrm$ to $\btrm$.  The term $\pchk{\atrm}$
discharges the guard on $\atrm$.  The evaluation rule allows
$\okeval{\arol}{\pchk{\pgrd{\brol}{\atrm}}}{\pblk{\atrm}}$ only if
$\okrsup{}{\arol}{\brol}$.

The context role may vary during evaluation: given context role
$\arol$, the term $\pmod{\arm}{\atrm}$ evaluates $\atrm$ with context
role $\arm[\arol]$.  Thus, when $\pmoddn{\brol}{\atrm}$ is evaluated
with context role $\arol$, $\atrm$ is evaluated with context role
$\arol\rless\brol$.  A context may protect itself from a term by
placing the use of the term in such a restricted context.  (The syntax
enforces a stack discipline on role modifiers.)  By combining
upwards and downwards modifiers, code may assume any role and thus
circumvent an intended policy.  We address this issue in
\autoref{sec:usercode}.

These constructs are sufficient to allow protection for both terms
and contexts: terms can be protected from contexts using guards, and
contexts can be protected from terms using (restrictive) role modifiers.

%
\subsection{Syntax}
\label{sec:syntax}

Let $\avar,\bvar,\cvar,\afun,\bfun$ range over variable names, and let
$\abase$ range over base values.  Our presentation is abstract with
respect to base values; we use the types $\tstring$, $\tint$ and
$\tunit$ (with value $\punit$) in examples.  We use the standard
encodings of booleans and pairs (see \autoref{ex:booleans}).  The
syntax of values and terms are as follows.

\smallskip
\begin{displaytab}{}
  \begin{math}
    \begin{aligned}&
      \aval,\bval,\cval\BNFDEF &\qquad&
      \atrm,\btrm,\ctrm\BNFDEF &\qquad&
      \text{Values; Terms}
      \\[-.5ex]&
      \qquad\abase\BNFSEP\avar&&
      \qquad\aval&&
      \qquad\text{Base Value}
      \\[-.5ex]&
      \qquad\pabs{\avar}{\atrm}&&
      \qquad\papp{\atrm}{\btrm}\BNFSEP\pfix{\atrm}&&
      \qquad\text{Abstraction}
      \\[-.5ex]&
      \qquad\pgrd{\arol}{\atrm}&&
      \qquad\pchk{\atrm}&&
      \qquad\text{Guard}
      \\[-.5ex]&
      \qquad\pblk{\atrm}&&
      \qquad\plet{\avar}{\atrm}\btrm&&
      \qquad\text{Computation}
      \\[-.5ex]&
      \qquad&&
      \qquad\pmod{\arm}{\atrm}&&
      \qquad\text{Role Modifier}
    \end{aligned}
  \end{math}
\end{displaytab}
\begin{notation}
  In examples, we write $\pmodeq{\arol}{\atrm}$ to abbreviate
  $\pmoddn{\rleast}{\pmodup{\arol}{\atrm}}$, which  executes $\atrm$
  at exactly role $\arol$.

  The variable $\avar$ is bound in the value ``$\pabs{\avar}{\atrm}$''
  (with scope $\atrm$) and in the term ``$\plet{\avar}{\atrm}\btrm$''
  (with scope $\btrm$).
  If $\avar$ does not appear free in $\atrm$, we abbreviate
  ``$\pabs{\avar}{\atrm}$'' as ``$\pabs{}{\atrm}$''.  Similarly, if
  $\avar$ does not appear free in $\btrm$, we abbreviate
  ``$\plet{\avar}{\atrm}\btrm$'' as ``$\plet{}{\atrm}\btrm$''.
  We identify syntax up to renaming of bound variables and write
  $\subst{\btrm}{\avar}{\atrm}$ for the capture-avoiding substitution
  of $\atrm$ for $\avar$ in $\btrm$.\qed
\end{notation}

In the presentation of the syntax above, we have paired the
constructors on values on the left with the destructors on
computations on the right.  For example, the monadic metalanguage
distinguishes $2$ from $\pblk{2}$ and $\pblk{1\OP+1}$: the former is
an integer, whereas the latter are computations that, when bound,
produce an integer.  The computation value $\pblk{\atrm}$ must be
discharged in a binding context --- see the reduction rule for let,
below.  Similarly, the function value $\pabs{\avar}{\atrm}$ must be
discharged by application; in the reduction semantics that follows,
evaluation proceeds in an application till the term in function
position reduces to a lambda abstraction.  $\pgrd{\arol}{\atrm}$
constructs a guarded value; the associated destructor is $\pchk$.

The monadic metalanguage distinguishes computations from the values
they produce and treats computations as first class entities.  (Any
term may be treated as a value via the unit constructor
$\pblk{\atrm}$.)  Both application and the let construct result in
computations; however, the way that they handle their arguments is
different.  The application
``$\papp{\pabsp{\avar}{\btrm}}{\pblk{\atrm}}$'' results in
$\subst{\btrm}{\avar}{\pblk{\atrm}}$, whereas the binding
``$\plet{\avar}{\pblk{\atrm}}\atrm$'' results in
$\subst{\btrm}{\avar}{\atrm}$.

%
\subsection{Evaluation and role error}
\label{sec:eval}

The small-step evaluation relation $\okeval{\arol}{\atrm}{\atrm'}$ is
defined inductively by the following reduction and context rules.
\smallskip
\begin{displaytab}{}
  \begin{math}
    \begin{aligned}&
      \linfer{r-app}{
      }{
        \okeval{\arol}{\papp{\pabsp{\avar}{\atrm}}{\btrm}}{\subst{\atrm}{\avar}{\btrm}}
      }
      &\qquad&
      \linfer{c-app}{
        \okeval{\arol}{\atrm}{\atrm'}
      }{
        \okeval{\arol}{\papp{\atrm}{\btrm}}{\papp{\atrm'}{\btrm}}
      }
      \\&
      \linfer{r-fix}{
      }{
        \okeval{\arol}{\pfix{\pabsp{\avar}{\atrm}}}{\subst{\atrm}{\avar}{\pfix{\pabsp{\avar}{\atrm}}}}
      }
      &\qquad&
      \linfer{c-fix}{
        \okeval{\arol}{\atrm}{\atrm'}
      }{
        \okeval{\arol}{\pfix{\atrm}}{\pfix{\atrm'}}
      }
      \\&
      \linferSIDE{r-chk}{
      }{
        \okeval{\arol}{\pchk{\pgrd{\brol}{\atrm}}}{\pblk{\atrm}}
      }{
        \okrsup{}{\arol}{\brol}
      }
      &&
      \linfer{c-chk}{
        \okeval{\arol}{\atrm}{\atrm'}
      }{
        \okeval{\arol}{\pchk{\atrm}}{\pchk{\atrm'}}
      }
      \\&
      \linfer{r-bind}{
      }{
        \okeval{\arol}{\plet{\avar}{\pblk{\atrm}}\btrm}{\subst{\btrm}{\avar}{\atrm}}
      }
      &&
      \linfer{c-bind}{
        \okeval{\arol}{\atrm}{\atrm'}
      }{
        \okeval{\arol}{\plet{\avar}{\atrm}\btrm}{\plet{\avar}{\atrm'}\btrm}
      }
      \\&
      \linfer{r-mod}{
      }{
        \okeval{\arol}{\pmod{\arm}{\aval}}{\aval}
      }
      &&
      \linfer{c-mod}{
        \okeval{\arm[\arol]}{\atrm}{\atrm'}
      }{
        \okeval{\arol}{\pmod{\arm}{\atrm}}{\pmod{\arm}{\atrm'}}
      }
    \end{aligned}
  \end{math}
\end{displaytab}

The rules \RN{r/c-app} for application, \RN{r/c-fix} for fixed points
and \RN{r/c-bind} for $\PF{let}$ are standard.  \RN{r-chk} ensures
that the context role is sufficient before discharging the relevant
guard.  \RN{c-mod} modifies the context role until the relevant term
is reduced to a value, at which point \RN{r-mod} discards the
modifier.

The evaluation semantics is designed to ensure a role-monotonicity
property.  Increasing the available role-context cannot invalidate
transitions, it can only enable more evolution.
\begin{lemma}
  \label{result:rednhigher}
  If $\okeval{\brol}{\atrm}{\atrm'}$
  and $\okrsup{}{\arol}{\brol}$
  then $\okeval{\arol}{\atrm}{\atrm'}$.
  \qed
\end{lemma}
\begin{proof}(Sketch)
  The context role is used only in \RN{r-chk}.  Result follows by
  induction on the evaluation judgement.
\end{proof}

Via a series of consecutive small steps, the final value for the
program can be determined.  Successful termination is written
$\okcvg{\arol}{\atrm}{\aval}$ which indicates that $\arol$ is
authorized to run the program $\atrm$ to completion, with result
$\aval$.  Viewed as a role-indexed relation on terms, $\evals$ is
reflexive and transitive.
\begin{definition}
  \begin{inparaenum}[(a)]
  \item $\atrm_0$ \emph{evaluates to} $\atrm_n$ at $\arol$ (notation
    $\okevals{\arol}{\atrm_0}{\atrm_n}$) if there exist terms
    $\atrm_i$ such that $\okeval{\arol}{\atrm_{i}}{\atrm_{i+1}}$, for
    all $i$ ($0\leq i\leq n-1$).
  \item $\atrm$ \emph{diverges} at $\arol$ (notation
    $\okdvg{\arol}{\atrm}$) if there exist terms $\atrm_i$ such that
    $\okeval{\arol}{\atrm_{i}}{\atrm_{i+1}}$, for all
    $i\in\mathbb{N}$.
    \qed
  \end{inparaenum}
\end{definition}

Evaluation can fail because a term diverges, because a destructor is
given a value of the wrong shape, or because an inadequate role is
provided at some point in the computation.  We refer to the latter as
a \emph{role error} (notation $\okerr{\arol}{\atrm}$), defined
inductively as follows.

\smallskip
\begin{displaytab}{}
  \linfernotags
  \linferSIDE{err-chk}{
  }{
    \okerr{\arol}{\pchkgrd{\brol}{\atrm}}
  }{
    \notokrsup{}{\arol}{\brol}
  }
  \QQUAD
  \linfer{err-ctx-mod}{
    \okerr{\arm[\arol]}{\atrm}
  }{
    \okerr{\arol}{\pmod{\arm}{\atrm}}
  }
  \\[1ex]
  \linfernotags
  \linfer{err-ctx-app}{
    \okerr{\arol}{\atrm}
  }{
    \okerr{\arol}{\papp{\atrm}{\btrm}}
  }
  \QQUAD
  \linfer{err-ctx-fix}{
    \okerr{\arol}{\atrm}
  }{
    \okerr{\arol}{\pfix{\atrm}}
  }
  \QQUAD
  \linfer{err-ctx-let}{
    \okerr{\arol}{\atrm}
  }{
    \okerr{\arol}{\plet{\avar}{\atrm}\btrm}
  }
  \QQUAD
  \linfer{err-ctx-chk}{
    \okerr{\arol}{\atrm}
  }{
    \okerr{\arol}{\pchk{\atrm}}
  }
\end{displaytab}

\begin{example}
  \label{ex:babyselinux}
Recall from \autoref{sec:syntax} that $\pmodeq{\brol}{\atrm}$
abbreviates $\pmoddn{\rleast}{\pmodup{\brol}{\atrm}}$,
  \begin{sfmath}
    and define $\secheckb$ as follows\footnote{We do not
      address parametricity here; the brackets in the names
      $\secheckb$ and $\seatob$ are merely suggestive.}.
    \begin{align*}
      \secheckb&\eqdef
      \pchkgrd{\brol}{\punit}
    \end{align*}
$\secheckb$ is a computation that requires context role $\brol$ to
evaluate.  For example, $\pmoddn{\rnegate\brol}{\secheckb}$ produces
a role error in any context, since $\rmdn{\rnegate\brol}{}$
restricts any role-context to the negation of the role $\brol$.
 \qed
  \end{sfmath}
\end{example}

\begin{example}
\label{ex:from}
\begin{sfmath}
We now illustrate how terms can provide roles for themselves. 
%
Consider the following guarded function:
    \begin{align*}
      \seatob&\eqdef
      \pgrd{\arol}{\pabs{\bvar}{
          \pmodeq{\brol}{\bvar}
        }}
    \end{align*}
$\seatob$ is a guarded value that may only be discharged by $\arol$,
resulting in a function that runs any computation at $\brol$. Let 
$\secheckb\eqdef \pchk{\pgrd{\brol}{\punit}}$.  
No matter what the relationship is between $\arol$ and $\brol$, the
following evaluation succeeds: 
    \begin{align*}&
      \okevals{\arol}{
        \plet{\cvar}{\pchk{\seatob}}\papp{\cvar}{\secheckb}
      }{
        \pmodeq{\brol}{\secheckb}
      }\evals{
        \pblk{\punit}
      }
    \end{align*}
$\seatob$ is far too powerful to be useful.  After the $\arol$-guard
is discharged, the resulting function will run \emph{any} code at role
$\brol$.  One can provide specific code, of course, as in 
\begin{math}
  {\pabs{\bvar}{\pmodeq{\brol}{\atrm}}}
\end{math}.
Such functions are inherently dangerous and therefore it is desirable
constrain the way in which such functions are created.  The essential
idea is to attach suitable checks to a function such as 
\begin{math}
  \pabsabs{\bfun}{\bvar}{\pmodeq{\brol}{\papp{\bfun}{\bvar}}}
\end{math}, which takes a non-privileged function and runs it under
$\brol$.  There are a number of subtleties to consider in providing a
general purpose infrastructure to create terms with rights
amplification.  When should the guard be checked?  What functions
should be allowed to run, and in what context?  In
\autoref{ex:selinux}, we discuss the treatment of these issues using the Domain
and Type Enforcement access control mechanism.
\qed
  \end{sfmath}
\end{example}

%
%
%

\section{Typing}
\label{sec:typing}

We present two typing systems that control role errors in addition
to shape errors.

The first typing system determines a context role \emph{sufficient}
to avoid role errors; that is, with this role, there is no possible
execution path that causes a role error.  This system enables the
removal of unnecessary role-checks in a piece of code for a caller
at a sufficiently high role.

The second system determines a context role \emph{necessary} to
avoid role errors; that is, any role that does not dominate this role
will cause every execution path to result in a role error. Stated
differently, the second system calculates the role that is checked
and tested on every execution path and thus determines the amount of
protection that is enforced by the callee.

Technically, the two systems differ primarily in their notions of
subtyping.  In the absence of subtyping, the typing system determines
a context role that is both necessary and sufficient to execute a
term without role errors.

Because it clearly indicates the point at which computation is
performed, the monadic metalanguage is attractive for reasoning about
security properties, which we understand as computational effects.
The type $\tblk{}{\atyp}$ is the type of computations of type
$\atyp$.  We extend the computation type $\tblk{}{\atyp}$ to include
an effect that indicates the guards that are discharged during
evaluation of a term. Thus the term $\pchk{\pgrd{\arol}{1\OP+1}$}
has type $\tblk{\arol}{\tint}$ --- this type indicates that the
reduction of the term to a value (at type $\tint$) requires $\arol$.
Guarded values inhabit types of the form $\tgrd{\arol}{\atyp}$ ---
this type indicates the protection of $\arol$ around an underlying
value at type $\atyp$.  These may be discharged with a $\PF{check}$,
resulting in a term inhabiting the computation type
$\tblk{\arol}{\atyp}$.

The syntax of types is given below, with the constructors and
destructors at each type recalled from \autoref{sec:syntax}.

\smallskip
\begin{displaytab}{}
  \begin{math}
    \begin{aligned}&
      \atyp,\btyp\BNFDEF &\qquad&
      \aval,\bval,\cval\BNFDEF &\qquad&
      \atrm,\btrm,\ctrm\BNFDEF &\qquad&
      \text{Types; Values; Terms}
      \\[-.5ex]&
      \qquad\tbase&&
      \qquad\abase\BNFSEP\avar&&
      \qquad\aval&&
      \qquad\text{Base Value}
      \\[-.5ex]&
      \qquad\tabs{\atyp}{\btyp}&&
      \qquad\pabs{\avar}{\atrm}&&
      \qquad\papp{\atrm}{\btrm}\BNFSEP\pfix{\atrm}&&
      \qquad\text{Abstraction}
      \\[-.5ex]&
      \qquad\tgrd{\arol}{\atyp}&&
      \qquad\pgrd{\arol}{\atrm}&&
      \qquad\pchk{\atrm}&&
      \qquad\text{Guard}
      \\[-.5ex]&
      \qquad\tblk{\arol}{\atyp}&&
      \qquad\pblk{\atrm}&&
      \qquad\plet{\avar}{\atrm}\btrm&&
      \qquad\text{Computation}
      \\[-.5ex]&
      \qquad&&
      \qquad&&
      \qquad\pmod{\arm}{\atrm}&&
      \qquad\text{Role Modifier}
    \end{aligned}
  \end{math}
\end{displaytab}

\subsection{Subtyping}
\label{sec:subtyping}


The judgments of the subtyping and typing relations are indexed by
$\theab$ which ranges over $\set{\thea,\theb}$.
The subtyping relation for $\tblk{\arol}{\atyp}$ reflects the
difference between the two type systems.

If role $\arol$ suffices to enable a term to evaluate without role
errors, then any higher role context
also avoids role errors (using \autoref{result:rednhigher}). This explains the subtyping rule for the
first type system
---  in particular,
$\aoktsub{}{\tblk{\arol}{\atyp}}{\tblk{\rmost}{\atyp}}$, reflecting
the fact that the top role is sufficient to run any computation.

On the other hand, if a role $\arol$ of the role-context is checked
and tested on every execution path of a term, then so is any smaller
role.  This explains the subtyping rule for the first type system
---  in particular, $\boktsub{}{\tblk{\arol}{\atyp}}{\tblk{\rleast}{\atyp}}$, reflecting
the fact that the bottom role is vacuously checked in any
computation.

\smallskip
\begin{displaytab}{}
  \begin{math}
    \begin{aligned}&
      \linfer{}{
      }{
        \soktsub{}{\tbase}{\tbase}
      }
      &&
      \linferSIDE{}{
        \soktsub{}{\atyp}{\atyp'}
      }{
        \soktsub{}{\tgrd{\arol}{\atyp}}{\tgrd{\arol'}{\atyp'}}
      }{
        \textif \theab=\thea \textthen\okrsub{}{\arol\phantom{'}}{\arol'}
        \\
        \textif \theab=\theb \textthen\okrsup{}{\arol\phantom{'}}{\arol'}
      }
      \\&
      \linfer{}{
        \soktsub{}{\atyp'}{\atyp}
        \QUAD
        \soktsub{}{\btyp}{\btyp'}
      }{
        \soktsub{}{\tabs{\atyp}{\btyp}}{\tabs{\atyp'}{\btyp'}}
      }
      &&
      \linferSIDE{}{
        \soktsub{}{\atyp}{\atyp'}
      }{
        \soktsub{}{\tblk{\arol}{\atyp}}{\tblk{\arol'}{\atyp'}}
      }{
        \textif \theab=\thea \textthen\okrsub{}{\arol\phantom{'}}{\arol'}
        \\
        \textif \theab=\theb \textthen\okrsup{}{\arol\phantom{'}}{\arol'}
      }
    \end{aligned}
  \end{math}
\end{displaytab}
\begin{lemma}
  \label{result:subtyperefltrans}
  The relations $\soktsub{}{\atyp}{\btyp}$ are reflexive and transitive.
  \qed
\end{lemma}

%
\subsection{Type systems}
\label{sec:typingsystems}

Typing is defined using environments.
An \emph{environment},
\begin{displaymath}
  \aenv\BNFDEF\VARTO{\avar}{\atyp}
\end{displaymath}
is a finite partial map from variables to types.

As usual, there is one typing rule for each syntactic form plus the
rule \RN{t-sub} for subsumption, which allows the use of subtyping.
Upwards and downwards role modifiers have separate rules, discussed
below.  The typing rules for the two systems differ only in their
notion of subtyping and in the side condition on \RN{t-mod-dn}; we
discuss the latter in \autoref{ex:t-mod-dn}.

\smallskip
\begin{displaytab}{}
  \begin{math}
    \begin{aligned}&
      \linfer{t-base}{
      }{
        \soktrm{\aenv}{\abase}{\tbase}
      }
      \quad
      \linfer{t-var}{
      }{
        \soktrm{\aenv,\VAR{\avar}{\atyp},\aenv'}{\avar}{\atyp}
      }
      &&
      \linferSIDE{t-sub}{
        \soktrm{\aenv}{\atrm}{\atyp}
      }{
        \soktrm{\aenv}{\atrm}{\atyp'}
      }{
        \soktsub{\aenv}{\atyp}{\atyp'}
      }
      \\&
      \linferSIDE{t-abs}{
        \soktrm{\aenv,\VAR{\avar}{\atyp}}{\atrm}{\btyp}
      }{
        \soktrm{\aenv}{\pabs{\avar}{\atrm}}{\tabs{\atyp}{\btyp}}
      }{
        \avar\notin\dom{\aenv}
      }
      &&
      \linfer{t-app}{
        \soktrm{\aenv}{\atrm}{\tabs{\atyp}{\btyp}}
        \QUAD
        \soktrm{\aenv}{\btrm}{\atyp}
      }{
        \soktrm{\aenv}{\papp{\atrm}{\btrm}}{\btyp}
      }
      \qquad
      \linfer{t-fix}{
        \soktrm{\aenv}{\atrm}{\tabs{\atyp}{\atyp}}
      }{
        \soktrm{\aenv}{\pfix{\atrm}}{\atyp}
      }
      \\&
      \linfer{t-grd}{
        \soktrm{\aenv}{\atrm}{\atyp}
      }{
        \soktrm{\aenv}{\pgrd{\arol}{\atrm}}{\tgrd{\arol}{\atyp}}
      }
      &&
      \linfer{t-chk}{
        \soktrm{\aenv}{\atrm}{\tgrd{\arol}{\atyp}}
      }{
        \soktrm{\aenv}{\pchk{\atrm}}{\tblk{\arol}{\atyp}}
      }
      \\&
      \linfer{t-unit}{
        \soktrm{\aenv}{\atrm}{\atyp}
      }{
        \soktrm{\aenv}{\pblk{\atrm}}{\tblk{\rleast}{\atyp}}
      }
      &&
      \linferSIDE{t-bind}{
        \soktrm{\aenv}{\atrm}{\tblk{\arol}{\atyp}}
        \QUAD
        \soktrm{\aenv,\VAR{\avar}{\atyp}}{\btrm}{\tblk{\brol}{\btyp}}
      }{
        \soktrm{\aenv}{\plet{\avar}{\atrm}\btrm}{\tblk{\arol\rmore\brol}{\btyp}}
      }{
        \avar\notin\dom{\aenv}
      }
      \\&
      \linfer{t-mod-up}{
        \soktrm{\aenv}{\atrm}{\tblk{\brol}{\atyp}}
      }{
        \soktrm{\aenv}{\pmodup{\arol}{\atrm}}{\tblk{\brol\rminus\arol}{\atyp}}
      }
      &&
      \linferSIDE{t-mod-dn}{
        \soktrm{\aenv}{\atrm}{\tblk{\brol}{\atyp}}
      }{
        \soktrm{\aenv}{\pmoddn{\arol}{\atrm}}{\tblk{\brol}{\atyp}}
      }{
        \textif \theab=\thea \textthen
        \okrsup{}{\arol}{\brol}
      }
    \end{aligned}
  \end{math}
\end{displaytab}

The rules \RN{t-base}, \RN{t-var}, \RN{t-sub}, \RN{t-abs},
\RN{t-app} and \RN{t-fix} are standard.  For example, the identity function has the
expected typing,
\begin{math}
  \soktrm{}{
    \pabs{\avar}{\avar}
  }{
    \tabs{\atyp}{\atyp}
  },
\end{math}
for any $\atyp$.
Nonterminating computations can also be typed; for example,
\begin{math}
  \soktrm{}{
    \pfix{\pabsp{\avar}{\avar}}
  }{
    \atyp
  },
\end{math}
for any $\atyp$.

Any term may be injected into a computation type at the least role
using \RN{t-unit}.  Thus, in the light of the earlier discussion on
subtyping, if $\soktrm{}{\atrm}{\atyp}$ then, in the first system,
$\pblk{\atrm}$ inhabits $\tblk{\arol}{\atyp}$ for every role
$\arol$; in the second system, the term inhabits only type
$\tblk{\rleast}{\atyp}$, indicating that no checks are required to
successfully evaluate the value $\pblk{\atrm}$.

Computations may be combined using \RN{t-bind}\footnote{The
distinction between our system and
  dependency-based systems can be see in \RN{t-bind}, which in
  \textsc{dcc} \cite{1159839,292555,tse04} states that
  $\oktrm{}{\plet{\avar}{\atrm}\btrm}{\tblk{\brol}{\btyp}}$ if
  $\brol\rsup\arol$, where
  $\oktrm{}{\atrm}{\tblk{\arol}{\atyp}}$ and
  $\oktrm{\VAR{\avar}{\atyp}}{\btrm}{\tblk{\brol}{\btyp}}$.}.  If $\atrm$ inhabits
$\tblk{\arol}{\atyp}$ and $\btrm$ inhabits $\tblk{\brol}{\btyp}$,
then ``$\plet{}{\atrm}\btrm$'' inhabits
$\tblk{\arol\rmore\brol}{\btyp}$.  More generally, we can deduce:
\[\soktrm{}{
        \pabs{\avar}{\plet{\avar'}{\avar}\avar'}
      }{
        \tabs{\tblk{\arol}{\tblk{\brol}{\atyp}}}{\tblk{\arol\rmore\brol}{\atyp}}
      }
      \]
In the first type system, this rule is motivated by noting that the
role context $\arol\rmore\brol$ suffices to successfully avoid role
errors in the combined computation if $\arol$ (resp. $\brol$)
suffices for $\atrm$ (resp. $\btrm$).  For the second type system,
consider a role $\crol$ that is not bigger than $\arol\rmore\brol$
--- thus $\crol$ is not bigger than at least one of $\arol,\brol$.
If it is not greater than $\arol$, by assumption on typing of
$\atrm$, every computation path of $\atrm$ in role context $\crol$
leads to a role-error.  Similarly for $\brol$. Thus, in role context
$\crol$, every computation path in the combined computation leads to
a role error. Furthermore, using the earlier subtyping discussion,
the sequence also inhabits $\tblk{\rmost}{\btyp}$ in the first
system and $\tblk{\rleast}{\btyp}$ in the second.

The rule \RN{t-grd} types basic values with their protection level.
The higher-order version of $\pgrd{\arol}{}$ has the natural typing:
\[ \soktrm{}{
        \pabs{\avar}{\pgrd{\arol}{\avar}}
      }{
        \tabs{\atyp}{\tgrd{\arol}{\atyp}}
      }
      \]
Recall that in the transition relation, $\pchk{\pgrd{\arol}{\btrm}}$
checks the role context against $\arol$.   The typing rule
\RN{t-chk} mirrors this behavior by converting the protection level
of values into constraints on role contexts.   For example, we have
the typing:
\[\soktrm{}{
        \pabs{\avar}{\pchk{\avar}}
      }{
        \tabs{\tgrd{\arol}{\atyp}}{\tblk{\arol}{\atyp}}
      }
\]
In the special case of typing
$\soktrm{\aenv}{\pchk{\pgrd{\arol}{\btrm}}}{\tblk{\arol}{\atyp}}$,
we can further justify in the two systems as follows.  In terms of
the first type system, the role context passes this check if it is
at least $\arol$.  In terms of the second type system, any role
context that does not include $\arol$ will cause a role-error.


Role modifiers are treated by separate rules for upwards and
downwards modifiers.

The rule for \RN{t-mod-up} is justified for the first type system as
follows. Under assumption that $\brol$ suffices to evaluate $\atrm$
without role-errors, consider evaluation of $\pmodup{\arol}{\atrm}$
in role context $\brol\rminus\arol$. This term contributes $\arol$
to role context yielding $\arol \rmore(\brol\rminus\arol) = (\arol
\rmore \brol) \rless (\arol\rmore \rnegate{\arol}) = \brol$ for the
evaluation of $\atrm$.  For the second type system, assume that if a
role is not greater than $\brol$, then the evaluation of $\btrm$
leads to a role error.  Consider the evaluation of
$\pmodup{\arol}{\atrm}$ in a role context $\crol$ that does not
exceed $\brol\rminus\arol$.  Then, the evaluation of $\atrm$
proceeds in role context $\crol \rmore \arol$ which does not exceed
$\brol$ and hence causes a role error by assumption.

The rule for \RN{t-mod-dn} is justified for the first type system as
follows. Under assumption that $\brol$ suffices to evaluate $\atrm$
without role-errors, and $\arol$ is greater than $\brol$ consider
evaluation of $\pmoddn{\arol}{\atrm}$ in role context $\brol$. This
term alters role-context $\brol$ to $\brol \rless \arol = \brol$ for
the evaluation of $\atrm$, which suffices.    For the second type
system, assume that if a role is not greater than $\brol$, then the
evaluation of $\btrm$ leads to a role error. Consider the evaluation
of $\pmoddn{\arol}{\atrm}$ in a role context $\crol$ that does not
exceed $\brol$.  Then, $\crol\rless\arol$ certainly does not exceed
$\brol$ and so the evaluation of $\atrm$ causes a role error by
assumption.

\autoref{ex:t-mod-*} and \autoref{ex:t-mod-dn} discuss alternate
presentations for the rules of typing for the role modifiers.

In stating the results, we distinguish computations from other types.
\autoref{result:dominates} holds
trivially from the definitions.
\begin{definition}
  Role $\arol$ \emph{dominates} type $\atyp$ (notation $\dominates{\arol}{\atyp}$) if
  $\atyp$ is not a computation type, or
  $\atyp$ is a computation type $\tblk{\brol}{\btyp}$ and
  $\okrsup{}{\arol}{\brol}$.
  \qed
\end{definition}
\begin{lemma}
  \label{result:dominates}
  \begin{inparaenum}[(a)]
  \item
    If $\okrsup{}{\arol}{\brol}$ and $\dominates{\brol}{\atyp}$ then
    $\dominates{\arol}{\atyp}$.
  \item
    \label{result:first:subdominates}
    If $\aoktsub{}{\atyp}{\btyp}$ and $\dominates{\arol}{\btyp}$
    then $\dominates{\arol}{\atyp}$.
  \item
    \label{result:second:subdominates}
    If $\boktsub{}{\atyp}{\btyp}$ and $\dominates{\arol}{\atyp}$
    then $\dominates{\arol}{\btyp}$.
    \qed
  \end{inparaenum}
\end{lemma}

The following theorems formalize the guarantees provided by the two
systems.
The proofs may be found in \autoref{sec:proofs}.
\begin{theorem}
  \label{result:first:divergeorconverge}
  If $\aoktrm{}{\atrm}{\atyp}$ and
  $\dominates{\arol}{\atyp}$, then either $\okdvg{\arol}{\atrm}$ or
  $\okcvg{\arol}{\atrm}{\aval}$ for some $\aval$.
\end{theorem}
\begin{theorem}
  \label{result:second:divergeorerror}
  If $\boktrm{}{\atrm}{\atyp}$ and
  $\notdominates{\arol}{\atyp}$, then either $\okdvg{\arol}{\atrm}$ or
  there exists $\btrm$ such that $\okevals{\arol}{\atrm}{\btrm}$ and
  $\okerr{\arol}{\btrm}$.
\end{theorem}

For the first system, we have a standard type-safety theorem.  For the
second system, such a safety theorem does not hold; for example
$\boktrm{}{\pchkgrd{\rmost}{\punit}}{\tblk{\rmost}{\tunit}}$ and
$\okeval{\rmost}{\pchkgrd{\rmost}{\punit}}{\pblk{\punit}}$ but
$\notboktrm{}{\pblk{\punit}}{\tblk{\rmost}{\tunit}}$.  Instead
\autoref{result:second:divergeorerror} states that a term run with an
insufficient context role is guaranteed either to diverge or to
produce a role error.

%
\subsection{Simple examples}
\label{sec:simple}

\begin{example}
  \begin{sfmath}
    We illustrate combinators of the language with some simple functions.
    The identity function may be given its usual type:
    \begin{displaymath}
      \soktrm{}{
        \pabs{\avar}{\avar}
      }{
        \tabs{\atyp}{\atyp}
      }
    \end{displaymath}
    The unit of computation can be used to create a computation from
    any value:
    \begin{displaymath}
      \soktrm{}{
        \pabs{\avar}{\pblk{\avar}}
      }{
        \tabs{\atyp}{\tblk{\rleast}{\atyp}}
      }
    \end{displaymath}
    The let construct evaluates a computation.  In this following
    example, the result of the computation $\avar'$ must itself be a
    computation because it is returned as the result of the function:
    \begin{displaymath}
      \soktrm{}{
        \pabs{\avar}{\plet{\avar'}{\avar}\avar'}
      }{
        \tabs{\tblk{\arol}{\tblk{\brol}{\atyp}}}{\tblk{\arol\rmore\brol}{\atyp}}
      }
    \end{displaymath}
    The guard construct creates a guarded term:
    \begin{displaymath}
      \soktrm{}{
        \pabs{\avar}{\pgrd{\arol}{\avar}}
      }{
        \tabs{\atyp}{\tgrd{\arol}{\atyp}}
      }
    \end{displaymath}
    The check construct discharges a guard, resulting in a computation:
    \begin{displaymath}
      \soktrm{}{
        \pabs{\avar}{\pchk{\avar}}
      }{
        \tabs{\tgrd{\arol}{\atyp}}{\tblk{\arol}{\atyp}}
      }
    \end{displaymath}
    The upwards role modifier reduces the role required by a computation.
    \begin{displaymath}
      \soktrm{}{
        \pabs{\avar}{\pmodup{\brol}{\avar}}
      }{
        \tabs{\tblk{\arol}{\atyp}}{\tblk{\arol\rminus\brol}{\atyp}}
      }
    \end{displaymath}
    The first typing system requires that any computation performed in
    the context of a downward role modifier $\pmoddn{\brol}{}$ must
    not require more than role $\brol$:
    \begin{displaymath}
      \soktrm{}{
        \pabs{\avar}{\pmoddn{\brol}{\avar}}
      }{
        \tabs{\tblk{\arol}{\atyp}}{\tblk{\arol}{\atyp}}
      }
      \;\;\;\;\text{(where $\okrsup{}{\brol}{\arol}$ if $\theab=\thea$)}
    \end{displaymath}
    In the first type system, the last two judgments may be
    generalized as follows:
    \begin{displaymath}
      \aoktrm{}{
        \pabs{\avar}{\pmod{\arm}{\avar}}
      }{
        \tabs{\tblk{\arm[\arol]}{\atyp}}{\tblk{\arol}{\atyp}}
      }
    \end{displaymath}
    Thus a role modifier may be seen as transforming a computation
    that requires the modifier into one that does not. For further
    discussion see \autoref{ex:t-mod-*}.\qed
  \end{sfmath}
\end{example}

\begin{example}[Booleans]
  \label{ex:booleans}
  The Church Booleans,
  \begin{sfmath}
    $\tru\eqdef\pabs{t}{\pabs{f}{t}}$ and
    $\fls\eqdef\pabs{t}{\pabs{f}{f}}$,
  \end{sfmath}
  illustrate the use of subtyping.  In the two systems, these
  may be given the following types.
  \begin{align*}&
    \atbool\eqdef
    \tabs{\tblk{\arol}{\atyp}}{
      \tabs{\tblk{\brol}{\atyp}}{
        \tblk{\arol\rmore\brol}{\atyp}}}
    &
    \aoktrm{}{\tru,\fls}{\atbool}
    \\&
    \btbool\eqdef
    \tabs{\tblk{\arol}{\atyp}}{
      \tabs{\tblk{\brol}{\atyp}}{
        \tblk{\arol\rless\brol}{\atyp}}}
    &
    \boktrm{}{\tru,\fls}{\btbool}
  \end{align*}
  These types reflect the intuitions underlying the two type
  systems.  The first type system reflects a ``maximum over all paths'' typing,
  whereas the second reflects a ``minimum over all paths'' typing.
  The conditional may be interpreted using the following derived rules.
  \begin{align*}&
    \linfer{}{
      \aoktrm{\aenv}{\ctrm}{\atbool}
      \quad
      \aoktrm{\aenv}{\atrm}{\tblk{\arol}{\atyp}}
      \quad
      \aoktrm{\aenv}{\btrm}{\tblk{\brol}{\atyp}}
    }{
      \aoktrm{\aenv}{\pcond{\ctrm}{\atrm}{\btrm}}{\tblk{\arol\rmore\brol}{\atyp}}
    }
    \\&
    \linfer{}{
      \boktrm{\aenv}{\ctrm}{\btbool}
      \quad
      \boktrm{\aenv}{\atrm}{\tblk{\arol}{\atyp}}
      \quad
      \boktrm{\aenv}{\btrm}{\tblk{\brol}{\atyp}}
    }{
      \boktrm{\aenv}{\pcond{\ctrm}{\atrm}{\btrm}}{\tblk{\arol\rless\brol}{\atyp}}
    }
  \amsqed
  \end{align*}
\end{example}

\begin{example}[\RN{t-mod-dn}]
  \label{ex:t-mod-dn}
  The side condition on \RN{t-mod-dn} does not effect typability in
  second typing system, but may unnecessarily decrease the accuracy of
  the analysis, as can be seen from the following concrete example.

  Let $\atrm,\btrm$ be terms of type $\tblk{\brol}{\atyp}$.
  \begin{displaymath}
    \cinfer[\RN{t-mod-dn}]{
      \cinfer[\RN{t-sub}]{
        \soktrm{\aenv}{\atrm}{\tblk{\brol}{\atyp}}
      }{
        \soktrm{\aenv}{\atrm}{\tblk{\arol\rless\brol}{\atyp}}
      }
    }{
      \soktrm{\aenv}{\pmoddn{\arol}{\atrm}}{\tblk{\arol\rless\brol}{\atyp}}
    }
  \end{displaymath}
  With the side condition, the term
  $\plet{\avar}{\pmoddn{\arol}{\atrm}}\btrm$ would have to be given
  a type of the form $\tblk{\arol\rless\brol}{\atyp}$, even though
  both $\atrm$ and $\btrm$ have type $\tblk{\brol}{\atyp}$.
  Without the side condition, the ``better'' type
  $\tblk{\brol}{\atyp}$ may be given to the entire $\PF{let}$
  expression.
  \qed
\end{example}

\begin{example}[Alternative rule for role modifiers]
  \label{ex:t-mod-*}
  In the first typing system, \RN{t-mod-up} and \RN{t-mod-dn} may be
  replaced with the following rule, which we call \RN{t-mod-*}.
  \begin{displaymath}
    \linfer{}{
      \aoktrm{\aenv}{\atrm}{\tblk{\arm[\brol]}{\atyp}}
    }{
      \aoktrm{\aenv}{\pmod{\arm}{\atrm}}{\tblk{\brol}{\atyp}}
    }
  \end{displaymath}
  \def\crol{C}%
  \def\drol{\arol\rmore\crol}%
  Consider $\arm=\rmup{\arol}{}$.  Because
  $\crol\rsup(\drol)\rminus\arol$, the following are equivalent.
  \begin{displaymath}
    \cinfer[\RN{t-mod-*}]{
      \aoktrm{\aenv}{\atrm}{\tblk{\arol\rmore\crol}{\atyp}}
    }{
      \aoktrm{\aenv}{\pmodup{\arol}{\atrm}}{\tblk{\crol}{\atyp}}
    }
    \QQQQQQQUAD
    \cinfer[\RN{t-sub}]{
      \cinfer[\RN{t-mod-up}]{
        \aoktrm{\aenv}{\atrm}{\tblk{\drol}{\atyp}}
      }{
        \aoktrm{\aenv}{\pmodup{\arol}{\atrm}}{\tblk{(\drol)\rminus\arol}{\atyp}}
      }
    }{
      \aoktrm{\aenv}{\pmodup{\arol}{\atrm}}{\tblk{\crol}{\atyp}}
    }
  \end{displaymath}
  \def\crol{\drol\rminus\arol}%
  \def\drol{D}%
  Because $(\crol)\rmore\arol\rsup\drol$, the following are
  equivalent.
  \begin{displaymath}
    \cinfer[\RN{t-mod-*}]{
      \cinfer[\RN{t-sub}]{
        \aoktrm{\aenv}{\atrm}{\tblk{\drol}{\atyp}}
      }{
        \aoktrm{\aenv}{\atrm}{\tblk{(\crol)\rmore\arol}{\atyp}}
      }
    }{
      \aoktrm{\aenv}{\pmodup{\arol}{\atrm}}{\tblk{\crol}{\atyp}}
    }
    \QQQQQQQUAD
    \cinfer[\RN{t-mod-up}]{
      \aoktrm{\aenv}{\atrm}{\tblk{\drol}{\atyp}}
    }{
      \aoktrm{\aenv}{\pmodup{\arol}{\atrm}}{\tblk{\drol\rminus\arol}{\atyp}}
    }
  \end{displaymath}
  \def\crol{C}%
  \def\drol{\arol\rless\crol}%
  Consider $\arm=\rmdn{\arol}{}$.  Because
  $\arol\rsup\drol$ and $\crol\rsup\drol$, the following are equivalent.
  \begin{displaymath}
    \cinfer[\RN{t-mod-*}]{
      \aoktrm{\aenv}{\atrm}{\tblk{\arol\rless\crol}{\atyp}}
    }{
      \aoktrm{\aenv}{\pmoddn{\arol}{\atrm}}{\tblk{\crol}{\atyp}}
    }
    \QQQQQQQUAD
    \cinfer[\RN{t-sub}]{
      \cinfer[\RN{t-mod-dn}]{
        \aoktrm{\aenv}{\atrm}{\tblk{\drol}{\atyp}}
      }{
        \aoktrm{\aenv}{\pmoddn{\arol}{\atrm}}{\tblk{\drol}{\atyp}}
      }
    }{
      \aoktrm{\aenv}{\pmoddn{\arol}{\atrm}}{\tblk{\crol}{\atyp}}
    }
  \end{displaymath}
  \def\crol{\drol}%
  \def\drol{D}%
  Suppose $\arol\rsup\drol$.
  Then $\crol\rless\arol\rsup\crol$, and the following are
  equivalent.
  \begin{displaymath}
    \cinfer[\RN{t-mod-*}]{
      \cinfer[\RN{t-sub}]{
        \aoktrm{\aenv}{\atrm}{\tblk{\crol}{\atyp}}
      }{
        \aoktrm{\aenv}{\atrm}{\tblk{\crol\rless\arol}{\atyp}}
      }
    }{
      \aoktrm{\aenv}{\pmoddn{\arol}{\atrm}}{\tblk{\crol}{\atyp}}
    }
    \QQQQQQQUAD
    \cinfer[\RN{t-mod-dn}]{
      \aoktrm{\aenv}{\atrm}{\tblk{\drol}{\atyp}}
    }{
      \aoktrm{\aenv}{\pmoddn{\arol}{\atrm}}{\tblk{\drol}{\atyp}}
    }
  \amsqed
  \end{displaymath}
\end{example}
\begin{example}[A sublanguage]
  The following proper sublanguage is sufficient to encode the
  computational lambda calculus.  Here values and terms are disjoint,
  with values assigned value types $\atyp$ and terms assigned
  computation types $\tblk{\arol}{\atyp}$.
  \begin{align*}
    \atyp,\btyp&\BNFDEF
    \tbase\BNFSEP
    \tabs{\atyp}{\tblk{\arol}{\btyp}}\BNFSEP
    \tgrd{\arol}{\atyp}
    \\
    \aval,\bval,\cval&\BNFDEF
    \abase\BNFSEP\avar\BNFSEP
    \pabs{\avar}{\atrm}\BNFSEP
    \pgrd{\arol}{\aval}
    \\
    \atrm,\btrm,\ctrm&\BNFDEF
    \pblk{\aval}\BNFSEP
    \papp{\aval}{\bval}\BNFSEP
    \pfix{\aval}\BNFSEP
    \pchk{\aval}\BNFSEP
    \plet{\avar}{\atrm}\btrm\BNFSEP
    \pmod{\arm}{\atrm}
  \end{align*}
  Encoding the Church Booleans in this sublanguage is slightly more
  complicated than in \autoref{ex:booleans}; $\tru$ and $\fls$ must
  accept thunks of type $\tabs{\tunit}{\tblk{\arol}{\btyp}}$ rather
  than the simpler blocks of type $\tblk{\arol}{\btyp}$.

  Operations on base values that have no computational effect are
  placed in the language of values rather than the language of terms.
  The resulting terms may be simplified at any time without affecting
  the computation (e.g., $\pblk{1\OP+2 \mathrel{\OP=\OP=} 3}$ may be simplified to
  $\pblk{\tru}$).
  \qed
\end{example}

\begin{example}[Relation to conference version]
  \newcommand{\OLDtabs}[4]{\tabs{#3}{\tgrd{#1\triangleright#2}{#4}}}
  \newcommand{\OLDpabs}[4]{\pgrd{#1}{\pabs{#3}{#4}}}
  \newcommand{\OLDpapp}[3]{\rmdn{#1}{}\;\papp{#2}{#3}}

  The language presented here is much simpler than that of the
  conference version of this
  paper~\cite{DBLP:conf/icalp/JagadeesanJPR06}.  In particular, the
  conference version collapsed guards and abstractions into a single
  form $\OLDpabs{\arol}{}{\avar}{\atrm}$ with types of the form
  $\OLDtabs{\arol}{\brol}{\atyp}{\btyp}$, which translates here as
  $\tgrdbig{\arol}{\tabs{\atyp}{\tblk{\brol}{\btyp}}}$: the immediate
  guard of the abstraction is $\arol$, whereas the effect of applying
  the abstraction is $\brol$.

  In addition, the conference version collapsed role modification and
  application: the application $\OLDpapp{\crol}{\aval}{\bval}$ first
  checked the guard of $\aval$, then performed the application in a
  context modified by $\rmdn{\crol}{}$.  In the current presentation,
  this translates as
  ``$\plet{\avar}{\pchk{\aval}}\pmoddn{\crol}{\papp{\avar}{\bval}}$.''
  \qed
\end{example}

%
%
%
\newcommand*{\cpsefliptype}[1]{\overline{#1}}
\newcommand*{\cpsedomtrans}[3]{\RFUNTHREE{domtrans}{#1}{#2}{#3}}
\newcommand*{\cpseassign}[1]{\RFUNONE{assign}{#1}}
\newcommand{\eqstr}{\OP{==}}
\newcommand{\radmin}{\PF{Admin}}
\newcommand{\ralice}{\PF{Alice}}
\newcommand{\rbob}{\PF{Bob}}
\newcommand{\rcharlie}{\PF{Charlie}}
\newcommand{\rdaemon}{\PF{Daemon}}
\newcommand{\rdebug}{\PF{Debug}}
\newcommand{\rlogin}{\PF{Login}}
\newcommand{\ruser}{\PF{User}}
\newcommand{\xadmin}{\PF{AdminEXE}}
\newcommand{\xlogin}{\PF{LoginEXE}}
\newcommand{\xuser}{\PF{UserEXE}}
\newcommand{\xrol}{\erol}

\newcommand{\fsfilea}{\pstring{file1}}
\newcommand{\fsfileb}{\pstring{file2}}
\newcommand{\fsfileacontents}{\pstring{data1}}
\newcommand{\fsfilebcontents}{\pstring{data2}}
\newcommand{\fsfilenotfound}{\pstring{error: file not found}}
\newcommand{\fsname}{\PF{name}}
\newcommand{\fswebserver}{\PF{webserver}}
\newcommand{\fsfilesystem}{\PF{filesystem}}

\newcommand{\sepay}{g}
\newcommand{\sedtefun}{f}
\newcommand{\segrd}{z}
\newcommand{\searg}{x}
\newcommand{\sepwd}{\PF{pwd}}
\newcommand{\secmd}{\PF{cmd}}
\newcommand{\seassignx}{\PF{assignE}}
\newcommand{\seassignl}{\PF{assignXLogin}}
\newcommand{\seassignu}{\PF{assignXUser}}
\newcommand{\setrans}[3]{\RFUNTHREE{dt}{#1}{#2}{#3}}
\newcommand{\setransab}{\PF{dtA\!ToB}}
\newcommand{\setransdl}{\PF{dtDaemon\!ToLogin}}
\newcommand{\setranslu}{\PF{dtLogin\!ToUser}}
\newcommand{\selogin}{\PF{login}}
\newcommand{\seshell}{\PF{shell}}
\newcommand{\seassign}[1]{\RFUNONE{assign}{#1}}
\newcommand{\serun}[1]{\RFUNONE{run}{#1}}
\newcommand{\tdte}[3]{#1\xrightarrow{#2}{#3}}
\newcommand{\tdtetight}[3]{#1\mathord{\xrightarrow{#2}}{#3}}
\newcommand{\seusercode}[3]{\PF{Func}(#1, #2, #3)}
\newcommand{\seexetype}[4]{\PF{FuncDTEType}(#1, #2, #3, #4)}

\newcommand{\tdtep}[3]{\PBR{\tdte{#1}{#2}{#3}}}
\newcommand{\tdtetightp}[3]{\PBR{\tdtetight{#1}{#2}{#3}}}

                                %
\section{Examples}
\label{sec:examples}

In this section we assume nullary role constructors for {user
  roles}, 
such as $\ralice$, $\rbob$, $\rcharlie$, $\radmin$, and $\rdaemon$.

\begin{example}[ACLs]
  \label{ex:filesystem}
  \begin{sfmath}
    Consider a read-only filesystem protected by Access Control Lists (ACLs).
    One can model such a system as:
    \renewcommand{\arraycolsep}{2pt}
    \renewcommand{\pif}[2]{\pcond{#1}{#2}{}}
    \renewcommand{\pelseif}[2]{\KW{else}&\pcond{#1}{#2}{}}
    \renewcommand{\pelse}[1]{\KW{else}\ &#1}
    \begin{displaymath}
      \fsfilesystem \eqdef
      \begin{array}[t]{@{}rl}
        \pabs{\fsname}{&
          \pif{\fsname\eqstr\fsfilea}{\pchkgrd{\radmin}{\fsfileacontents}} \\
          \pelseif{\fsname\eqstr\fsfileb}{\pchkgrd{\ralice\rless\rbob}{\fsfilebcontents}} \\
          \pelse{\pblk{\fsfilenotfound}}}
      \end{array}
    \end{displaymath}
    If $\radmin \geq \ralice\rless\rbob$ then code
    running in the $\radmin$ role can access both files:
    \begin{displaymath}
      \begin{array}{@{}l}
        \oke{\radmin}
        \papp{\fsfilesystem}{\fsfilea}
        \evals
        \pchkgrd{\radmin}{\fsfileacontents}
        \evals
        \pblk{\fsfileacontents}
        \\
        \oke{\radmin}
        \papp{\fsfilesystem}{\fsfileb}
        \evals
        \pchkgrd{\ralice\rless\rbob}{\fsfilebcontents}
        \evals
        \pblk{\fsfilebcontents}
      \end{array}
    \end{displaymath}%
    If $\ralice \not\geq \radmin$ then code running as $\ralice$
    cannot access the first file but can access the second:
    \begin{displaymath}
      \begin{array}{@{}l}
        \oke{\ralice}
        \papp{\fsfilesystem}{\fsfilea}
        \evals
        \pchkgrd{\radmin}{\fsfileacontents}
        \okerr{}{}
        \\
        \oke{\ralice}
        \papp{\fsfilesystem}{\fsfileb}
        \evals
        \pchkgrd{\ralice\rless\rbob}{\fsfilebcontents}
        \evals
        \pblk{\fsfilebcontents}
      \end{array}
    \end{displaymath}%
    Finally, if $\rcharlie \not\geq \ralice\rless\rbob$ then code
    running as $\rcharlie$ cannot access either file:
    \begin{displaymath}
      \begin{array}[b]{@{}l}
        \oke{\rcharlie}
        \papp{\fsfilesystem}{\fsfilea}
        \evals
        \pchkgrd{\radmin}{\fsfileacontents}
        \okerr{}{}
        \\
        \oke{\rcharlie}
        \papp{\fsfilesystem}{\fsfileb}
        \evals
        \pchkgrd{\ralice\rless\rbob}{\fsfilebcontents}
        \okerr{}{}
      \end{array}
    \end{displaymath}%
    
    The filesystem code can be assigned the following type, meaning
    that a caller must possess a role from each of the ACLs in order
    to guarantee that access checks will not fail.  If, in addition,
    $\radmin \geq \ralice\rless\rbob$ then the final role is equal to
    $\radmin$.
    \begin{displaymath}
      \aoktrm{}{\fsfilesystem}{\tabs{\tstring}{
          \tblk{\radmin \rmore (\ralice\rless\rbob) \rmore \rleast}{\tstring}}}
    \end{displaymath}
    \noindent
    In the above type, the final role $\rleast$ arises from the
    ``unknown file'' branch that does not require an access check.
    The lack of an access check explains the weaker $\bdash$ type:
    \begin{displaymath}
      \boktrm{}{\fsfilesystem}{\tabs{\tstring}{
          \tblk{\radmin \rless (\ralice\rless\rbob) \rless \rleast}{\tstring}}}
    \end{displaymath}
    This type indicates that $\fsfilesystem$ has the potential to expose
    some information to unprivileged callers with role $\radmin \rless
    (\ralice\rless\rbob) \rless \rleast \req \rleast$, perhaps causing the code to be
    flagged for security review.
    \qed
  \end{sfmath}
\end{example}
\begin{example}[Web server]
  \begin{sfmath}
    Consider a web server that provides remote access to the
    filesystem described above.
    The web server can use the role assigned to a caller to access
    the filesystem (unless the web server's caller withholds its role).
    To prevent an attacker determining the non-existence of files via
    the web server, the web server  fails when an attempt is made to
    access an unknown file unless the $\rdebug$ role is activated.
    \renewcommand{\arraycolsep}{2pt}
    \renewcommand{\pif}[2]{\pcond{#1}{#2}{}}
    \renewcommand{\pelseif}[2]{\KW{else}&\pcond{#1}{#2}{}}
    \renewcommand{\pelse}[1]{\KW{else}\ &#1}
    \begin{displaymath}
      \fswebserver \eqdef
      \begin{array}[t]{@{}rl}
        \pabs{\fsname}{&
          \pif{\fsname\eqstr\fsfilea}{\papp{\fsfilesystem}{\fsname}} \\
          \pelseif{\fsname\eqstr\fsfileb}{\papp{\fsfilesystem}{\fsname}} \\
          \pelse{\pchkgrd{\rdebug}{\fsfilenotfound}}}
      \end{array}
    \end{displaymath}

    For example, code running as Alice can access $\fsfileb$ via the web
    server:
    \begin{displaymath}
      \oke{\ralice}
      \papp{\fswebserver}{\fsfileb}
      \evals
      \papp{\fsfilesystem}{\fsfileb}
      \evals
      \pblk{\fsfilebcontents}
    \end{displaymath}
    
    The access check in the web server does prevent the ``file not
    found'' error message leaking unless the $\rdebug$ role is active,
    but, unfortunately, it is not possible to assign a role strictly
    greater than $\rleast$ to the web server using the second type
    system.  The $\fsfilesystem$ type does not record the different
    roles that must be checked depending upon the filename argument.
    \begin{align*}&
      \boktrm{}{\fswebserver}{\tabs{\tstring}{
          \tblk{\radmin \rless (\ralice\rless\rbob) \rless \rleast}{\tstring}}}
      &&\text{(derivable)}
      \\&
      \not\boktrm{}{\fswebserver}{\tabs{\tstring}{
          \tblk{\radmin \rless (\ralice\rless\rbob) \rless \rdebug}{\tstring}}}
      &&\text{(not derivable)}
      \amsqed
    \end{align*}
  \end{sfmath}
\end{example}

\autoref{ex:selinux} illustrates how the \emph{Domain-Type
  Enforcement} (\DTE) access control mechanism
\cite{domain-type:boebert85,dte:confining-root-programs}, found in
Security-Enhanced Linux (\SELinux) \cite{selinux:smalley01}, can be
modelled in \lrbac{}.  Further discussion of the relationship
between \RBAC\ and \DTE\ can be found in
\cite{rbac-book,implementing-rbac:hoffman}.
\begin{example}[Domain-Type Enforcement]
  \label{ex:selinux}
  \begin{sfmath}
    The \DTE\ access control mechanism grants or denies access
    requests according to the current \emph{domain} of running code.
    The current domain changes as new programs are executed, and
    transitions between domains are restricted in order to allow, and
    also force, code to run with an appropriate domain.  The
    restrictions upon domain transitions are based upon a \DTE\ type
    associated with each program to execute.  For example, the \DTE\ 
    policy in \cite{dte:confining-root-programs} only permits
    transitions from a domain for daemon processes to a domain for
    login processes when executing the login program, because code
    running in the login domain is highly privileged.  This effect is
    achieved by allowing transitions from the daemon domain to the
    login domain only upon execution of programs associated with a
    particular \DTE\ type, and that \DTE\ type is assigned only to the
    login program.
    
    The essence of \DTE\ can be captured in \lrbac{}, using roles to
    model both domains and \DTE\ types, and the context role to model
    the current domain of a system.  We start by building upon the
    code fragment
    \begin{math}
      \pabsabs{\bfun}{\bvar}{\pmodeq{\brol}{\papp{\bfun}{\bvar}}}
    \end{math},
    discussed in \autoref{ex:from}, that allows a function checking
    role $\brol$ to be executed in the context of code running at a
    different role.  We have the typing (for emphasis we use extra
    parentheses that are not strictly necessary given the usual right
    associativity for the function type constructor):
    \begin{displaymath}
      \soktrm{}{
        \pabsabs{\bfun}{\bvar}{\pmodeq{\brol}{\papp{\bfun}{\bvar}}}}{
        \tabs{
          (\tabs{\atyp}{\tblk{\brol}{\btyp}})}{
          (\tabs{\atyp}{\tblk{\rleast}{\btyp}})}
        }
    \end{displaymath}
    To aid readability, and fixing types $\atyp$ and $\btyp$ for the
    remainder of this example, define:
    \begin{displaymath}
      \cpsefliptype{\ctyp} \eqdef
      \tabs{\ctyp}{(\tabs{\atyp}{\tblk{\rleast}{\btyp}})}
    \end{displaymath}
    So that the previous typing becomes:
    \begin{displaymath}
      \soktrm{}{
        \pabsabs{\bfun}{\bvar}{\pmodeq{\brol}{\papp{\bfun}{\bvar}}}}{
        \cpsefliptype{\tabs{\atyp}{\tblk{\brol}{\btyp}}}
        }
    \end{displaymath}
    To restrict the use of the privileged function
    \begin{math}
      \pabsabs{\bfun}{\bvar}{\pmodeq{\brol}{\papp{\bfun}{\bvar}}}
    \end{math}, 
    it can be guarded by a role $\erol$ acting as a \DTE\ type, where
    the association of the \DTE\ type $\erol$ with a function is
    modelled in the sequel by code that can activate role $\erol$.
    The guarded function can be typed as:
    \begin{displaymath}
      \soktrm{}{
        \pgrd{\erol}{\pabsabs{\bfun}{\bvar}{\pmodeq{\brol}{\papp{\bfun}{\bvar}}}}}{
        \tgrd{\erol}{\cpsefliptype{\tabs{\atyp}{\tblk{\brol}{\btyp}}}}
        }
    \end{displaymath}
    We now define a function $\cpsedomtrans{\arol}{\erol}{\brol}$ for
    a domain transition from domain (role) $\arol$ to domain (role)
    $\brol$ upon execution of a function associated with \DTE\ type
    (also a role) $\erol$.  The function first verifies that the
    context role dominates $\arol$, and then permits use of the
    privileged function
    \begin{math}
      \pabsabs{\bfun}{\bvar}{\pmodeq{\brol}{\papp{\bfun}{\bvar}}}
    \end{math}
    by code that can activate role $\erol$.  The function
    $\cpsedomtrans{\arol}{\erol}{\brol}$ is defined by:
    \begin{displaymath}
      \cpsedomtrans{\arol}{\erol}{\brol} \eqdef
      \pabsabs{\afun}{\avar}{
        \pchk{\pgrd{\arol}{\punit}} 
        \SEMI
        \pappapp{\afun}{
          \pgrd{\erol}{\pabsabs{\bfun}{\bvar}{\pmodeq{\brol}{\papp{\bfun}{\bvar}}}}}{
          \avar}
      }
    \end{displaymath}
    We have the typing:
    \begin{displaymath}
      \soktrm{}{
        \cpsedomtrans{\arol}{\erol}{\brol}
      }{
        \tabs{
          \cpsefliptype{
            \tgrd{
              \erol}{
              \cpsefliptype{
                \tabs{
                  \atyp}{
                  \tblk{\brol}{\btyp}
                }
              }
            }
          }
        }{
          (\tabs{
            \atyp}{
            \tblk{\arol}{\btyp}
          })
        }
      }
    \end{displaymath}
    The above type shows that $\cpsedomtrans{\arol}{\erol}{\brol}$ can
    be used to turn a function checking role $\brol$ into a function
    checking role $\arol$, but only when the role $\erol$ is
    available---in contrast to the type
    \begin{math}
      \tabs{
        \cpsefliptype{
          {
            \cpsefliptype{
              (\tabs{
                \atyp}{
                \tblk{\brol}{\btyp}
              })
            }
          }
        }
      }{
        (\tabs{
          \atyp}{
          \tblk{\arol}{\btyp}
        })
      }
    \end{math}
    that does not require $\erol$.
    
    In order to make use of $\cpsedomtrans{\arol}{\erol}{\brol}$, we
    must also consider code that can activate $\erol$.  We define a
    function $\cpseassign{\erol}$ that takes a function $\afun$ and
    activates $\erol$ in order to access the privileged code
    \begin{math}
      \pabsabs{\bfun}{\bvar}{\pmodeq{\brol}{\papp{\bfun}{\bvar}}}
    \end{math}
    from $\cpsedomtrans{\arol}{\erol}{\brol}$.  The function
    $\cpseassign{\erol}$ is defined by:
    \begin{displaymath}
      \cpseassign{\erol} \eqdef
      \pabsabs{\afun}{\avar}{
        \pabs{\bvar}{
          \plet{\bfun}{\pmodeq{\erol}{\pchk{\avar}}}
          \pappapp{\bfun}{\afun}{\bvar}
        }
      }
    \end{displaymath}
    And we have the typing:
    \begin{displaymath}
      \soktrm{}{\cpseassign{\erol}}{
        \tabs{
          (\tabs{
            \atyp}{
            \tblk{\brol}{\btyp}
          })
        }{
          \cpsefliptype{
            \tgrd{
              \erol}{
              \cpsefliptype{
                \tabs{
                  \atyp}{
                  \tblk{\brol}{\btyp}
                }
              }
            }
          }
        }
      }
    \end{displaymath}
    Therefore the functional composition of $\cpseassign{\erol}$ and
    $\cpsedomtrans{\arol}{\erol}{\brol}$ has type:
    \begin{displaymath}
      \tabs{
        (\tabs{
          \atyp}{
          \tblk{\brol}{\btyp}
        })
      }{
        (\tabs{
          \atyp}{
          \tblk{\arol}{\btyp}
        })
      }
    \end{displaymath}
    To show that in the presence of both $\cpseassign{\erol}$ and
    $\cpsedomtrans{\arol}{\erol}{\brol}$, code running with context
    $\arol$ can execute code checking for role context $\brol$, we
    consider the following reductions in role context $\arol$, where
    we take \newcommand{\xxx}{\mathcal{F}} 
    \begin{math}
      \xxx \eqdef {\pabs{\cvar}{\pchk{\pgrd{\brol}{\punit}}}}
    \end{math}
    and underline terms to indicate the redex:
    \begin{align*}
      &
      \pappapp
      {\underline{\cpsedomtrans{\arol}{\erol}{\brol}}}
      {\PBR{\papp{\cpseassign{\erol}}{\xxx}}}
      {\punit}
      \\=\;&
      \pappapp
      {\PBR{\underline{\pabsabs{\afun}{\avar}{
              \pchk{\pgrd{\arol}{\punit}} 
              \SEMI
              \pappapp{\afun}{
                \PBR{\pgrd{\erol}{\pabsabs{\bfun}{\bvar}{\pmodeq{\brol}{\papp{\bfun}{\bvar}}}}}}{
                \avar}
            }}}}
      {\underline{\PBR{\papp{\cpseassign{\erol}}{\xxx}}}}
      {\punit}
      \\\eval\;&
      \papp
      {\PBR{\underline{\pabs{\avar}{
              \pchk{\pgrd{\arol}{\punit}} 
              \SEMI
              \pappapp{\PBR{\papp{\cpseassign{\erol}}{\xxx}}}{
                \PBR{\pgrd{\erol}{\pabsabs{\bfun}{\bvar}{\pmodeq{\brol}{\papp{\bfun}{\bvar}}}}}}{
                \avar}
            }}}}
      {\underline{\punit}}
      \\\eval\;&
      \underline{\pchk{\pgrd{\arol}{\punit}}}
      \SEMI
      \pappapp{\PBR{\papp{\cpseassign{\erol}}{\xxx}}}{
        \PBR{\pgrd{\erol}{\pabsabs{\bfun}{\bvar}{\pmodeq{\brol}{\papp{\bfun}{\bvar}}}}}}{
        \punit}
      \\\eval\;&
      \pappapp{\PBR{\papp{\underline{\cpseassign{\erol}}}{\xxx}}}{
        \PBR{\pgrd{\erol}{\pabsabs{\bfun}{\bvar}{\pmodeq{\brol}{\papp{\bfun}{\bvar}}}}}}{
        \punit}
      \\=\;&
      \pappapp{\bigPBR{\papp{
            \PBR{\underline{\pabsabs{\afun}{\avar}{
                \pabs{\bvar}{
                  \plet{\bfun}{\pmodeq{\erol}{\pchk{\avar}}}
                  \pappapp{\bfun}{\afun}{\bvar}
                }
              }}}}{\underline{\xxx}}}}{
        \PBR{\pgrd{\erol}{\pabsabs{\bfun}{\bvar}{\pmodeq{\brol}{\papp{\bfun}{\bvar}}}}}}{
        \punit}
      \\\eval\;&
      \pappapp{
        \PBR{\underline{\pabs{\avar}{
              \pabs{\bvar}{
                \plet{\bfun}{\pmodeq{\erol}{\pchk{\avar}}}
                \pappapp{\bfun}{\xxx}{\bvar}
              }
            }}}}{
        \PBR{\underline{\pgrd{\erol}{\pabsabs{\bfun}{\bvar}{\pmodeq{\brol}{\papp{\bfun}{\bvar}}}}}}}{
        \punit}
      \\\eval\;&
      \papp{
        \PBR{\underline{
          \pabs{\bvar}{
            \plet{\bfun}{\pmodeq{\erol}{\pchk{{\pgrd{\erol}{\pabsabs{\bfun}{\bvar}{\pmodeq{\brol}{\papp{\bfun}{\bvar}}}}}}}}
            \pappapp{\bfun}{\xxx}{\bvar}
          }
        }}}{
        \underline{\punit}}
      \\\eval\;&
      \plet{\bfun}{\pmodeq{\erol}{\underline{\pchk{{\pgrd{\erol}{\pabsabs{\bfun}{\bvar}{\pmodeq{\brol}{\papp{\bfun}{\bvar}}}}}}}}}
      \pappapp{\bfun}{\xxx}{\punit}
      \\\eval\;&
      \plet{\bfun}{\underline{\pmodeq{\erol}{\pblk{\pabsabs{\bfun}{\bvar}{\pmodeq{\brol}{\papp{\bfun}{\bvar}}}}}}}
      \pappapp{\bfun}{\xxx}{\punit}
      \\\evals\;&
      \underline{\plet{\bfun}{\pblk{\pabsabs{\bfun}{\bvar}{\pmodeq{\brol}{\papp{\bfun}{\bvar}}}}}
      \pappapp{\bfun}{\xxx}{\punit}}
      \\\eval\;&
      \pappapp{\PBR{\underline{\pabsabs{\bfun}{\bvar}{\pmodeq{\brol}{\papp{\bfun}{\bvar}}}}}}{\underline{\xxx}}{\punit}
      \\\eval\;&
      \papp{\PBR{\underline{\pabs{\bvar}{\pmodeq{\brol}{\papp{\xxx}{\bvar}}}}}}{\underline{\punit}}
      \\\eval\;&
      \pmodeq{\brol}{\papp{\underline{\xxx}}{\punit}}
      \\=\;&
      \pmodeq{\brol}{\papp{\PBR{\underline{\pabs{\cvar}{\pchk{\pgrd{\brol}{\punit}}}}}}{\underline{\punit}}}
      \\\eval\;&
      \pmodeq{\brol}{\underline{\pchk{\pgrd{\brol}{\punit}}}}
      \\\eval\;&
      \underline{\pmodeq{\brol}{\pblk{\punit}}}
      \\\evals\;&
      \pblk{\punit}
    \end{align*}
    
    The strength of \DTE\ lies in the ability to factor access control
    policies into two components: the set of permitted domain
    transitions and the assignment of \DTE\ types to code.  We
    illustrate this by adapting the aforementioned login example from
    \cite{dte:confining-root-programs} to $\lrbac$.  In this example,
    the \DTE\ mechanism is used to force every invocation of user code
    (running at role $\ruser$) from daemon code (running at role
    $\rdaemon$) to occur via trusted login code (running at role
    $\rlogin$).  This is achieved by providing domain transitions from
    $\rlogin$ to $\ruser$, and $\rdaemon$ to $\rlogin$, but no others.
    Moreover, code permitted to run at $\rlogin$ must be assigned
    \DTE\ type $\xlogin$, and similarly for $\ruser$ and $\xuser$.
    Thus a full program running daemon code $\atrm$ has the following
    form, where neither $\atrm$ nor $\btrm$ contain direct rights
    amplification.
    \allowdisplaybreaks
    \begin{displaymath}
      \begin{array}{l}
        \plet{\setranslu}{\cpsedomtrans{\rlogin}{\xuser}{\ruser}}{} \\
        \plet{\setransdl}{\cpsedomtrans{\rdaemon}{\xlogin}{\rlogin}}{} \\
        \plet{\seshell}{\papp{\cpseassign{\xuser}}{\pabsp{}{\atrm}}} \\
        \pletNOSEMI{\selogin}{%
          \papp{\cpseassign{\xlogin}}{%
            (\pabs{\sepwd}{
              \pcond{\sepwd\eqstr\text{"secret"}}{\papp{\papp{\setranslu}{\seshell}}{\punit}}{\!\dots}
              )\SEMI
            }}}
        \\
        \pmodeq{\rdaemon}{\btrm}
      \end{array}
    \end{displaymath}
    Because $\selogin$ provides the sole gateway to the role $\ruser$,
    the daemon code $\btrm$ must provide the correct password in order
    to execute the shell at $\ruser$ (in order to access resources
    that are available at role $\ruser$ but not at role $\rdaemon$).
    In addition, removal of the domain transition $\setransdl$ makes
    it impossible for the daemon code to execute any code at $\ruser$.
  \end{sfmath}
  \qed
\end{example}

\newcommand{\ucat}[1]{\RFUNONE{at}{#1}}
\newcommand{\ucpriv}{\PF{priv}}

%
\section{Controlling rights amplification}
\label{sec:usercode}

\begin{example}
\begin{sfmath}
  %
  Suppose that $\atrm$ contains no direct rights amplification, that
  is, no subterms of the form $\pmodup{\arol}{\cdot}$.  Then, in
  \begin{displaymath}
    \pletblk{\ucpriv}{\pabs{\avar}{\pmodup{\arol}{\papp{\aval}{\avar}}}}
    \pmoddn{\ruser}{\atrm}
  \end{displaymath}
  we may view $\aval$ as a \emph{Trusted Computing Base} (\TCB) --- a
  privileged function which may escalate rights --- and view $\atrm$ as
  restricted \emph{user code}.  The function $\ucpriv$ is an entry
  point to the \TCB\ which is accessible to user code; that is, user
  code is executed at the restricted role $\ruser$, and rights
  amplification may only occur through invocation of $\ucpriv$.

  Non-trivial programs have larger \TCB s with more entry points.  As
  the size of the \TCB\ grows, it becomes difficult to understand the
  security guarantees offered by a system when rights amplification is
  unconstrained, even if only in the \TCB.
  To manage this complexity, one may enforce a coding convention that
  requires rights increases be justified by earlier checks.  As an
  example, consider the following, where $\ramplify{}$ is a unary role
  constructor.
  \begin{displaymath}
    \begin{array}{@{}l}
      \pletblk{\ucat{\arol}}{
        \pabs{\afun}{
          \pchkgrd{\ramplify{\arol}}{
            \pabs{\avar}{
              \pmodup{\arol}{\papp{\afun}{\avar}}
            }}}}\\
      \plet{\ucpriv}{\papp{\ucat{\arol}}{\aval}}\\
      \pmoddn{\ruser}{\atrm}
    \end{array}
  \end{displaymath}
  In a context with role $\ramplify{\arol}$, this reduces (using
  \RN{r-bind}, \RN{r-app} and \RN{r-chk}) to
  \begin{displaymath}
    \begin{array}{@{}l}
      \pletblk{\ucpriv}{
        \pabs{\avar}{\pmodup{\arol}{\papp{\aval}{\avar}}}}
      \pmoddn{\ruser}{\atrm}
    \end{array}
  \end{displaymath}
  In a context without role $\ramplify{\arol}$, evaluation becomes
  stuck when attempting to execute \RN{r-chk}.
  The privileged function returned by $\ucat{\arol}$ (which
  performs rights amplification for $\arol$) is justified by the check
  for $\ramplify{\arol}$ on any caller of $\ucat{\arol}$.

  One may also wish to explicitly prohibit a term from direct
  amplification of some right $\brol$; with such a convention in
  place, this can be achieved using the role modifier
  $\rmdn{\ramplify{\brol}}{}$.
  \qed
\end{sfmath}
\end{example}

%
\label{sec:amplify}

One may formalize the preceding example by introducing the unary role
constructor $\ramplify{}$, where $\ramplify{\arol}$ stands for the
right to provide the role $\arol$ by storing $\rmup{\arol}{}$ in code.

We require that $\ramplify{}$ distribute over $\rmore$ and $\rless$
and obey the following absorption laws:
\begin{align*}
  \arol\rmore\ramplify{\arol}&\req\ramplify{\arol} &
  \arol\rless\ramplify{\arol}&\req\arol &
\end{align*}
Thus $\ramplify{\arol}\rsup\arol$ for any role $\arol$.

To distinguish justified use of role modifiers from unjustified use, we
augment the syntax with \emph{checked role modifiers}.
\begin{displaymath}
  \atrm,\btrm\BNFDEF\cdots\BNFSEP
  \pmod[\arol]{\arm}{\atrm}
\end{displaymath}
Whenever a check is performed on role $\atrm$ we \emph{mark} role
modifiers in the consequent to indicate that these modifiers have been
justified by a check.  Define the function $\fmark{\arol}{}$
homomorphically over all terms but for role modifiers:
\begin{align*}
  \fmark{\arol}{\pmod{\arm}{\atrm}} &= \pmod[\arol]{\arm}{\fmark{\arol}{\atrm}}
  \\
  \fmark{\arol}{\pmod[\brol]{\arm}{\atrm}} &= \pmod[\arol\rmore\brol]{\arm}{\fmark{\arol}{\atrm}}
\end{align*}
Modify the reduction rule for $\PF{check}$ as follows.
\begin{displaymath}
  \linferSIDE{}{
  }{
    \okeval{\arol}{\pchk{\pgrd{\brol}{\atrm}}}{\pblk{\fmark{\brol}{\atrm}}}
  }{
    \okrsup{}{\arol}{\brol}
  }
\end{displaymath}
Thus, the check in the example above will execute as follows.
\begin{displaymath}
  \okeval{\ramplify{\arol}}{
    \pchkgrd{\ramplify{\arol}}{
      \pabs{\avar}{
        \pmodup{\arol}{\papp{\afun}{\avar}}
      }}
  }{
    \pabs{\avar}{
      \pmodup[\ramplify{\arol}]{\arol}{\papp{\afun}{\avar}}
    }
  }
\end{displaymath}
In the residual, the abstraction contains a checked role modifier,
indicating that the role amplification has been provided by code that
had the right to do so.

We now define \emph{role modification errors} so that
$\pmodup[{\brol}]{\arol}{\atrm}$ produces an error if ${\brol}$ does
not dominate $\ramplify{\arol}$.
\begin{align*}&
  \linfer{}{
  }{
    \okamperr{}{\pmodup{\brol}{\atrm}}
  }
  \qquad
  \linferSIDE{}{
  }{
    \okamperr{}{\pmodup[\crol]{\brol}{\atrm}}
  }{
    \crol\not\rsup\ramplify{\brol}
  }
  \\&
  \linfer{}{
    \okamperr{}{\atrm}
  }{
    \okamperr{}{\papp{\atrm}{\btrm}}
  }
  \qquad
  \linfer{}{
    \okamperr{}{\atrm}
  }{
    \okamperr{}{\plet{\avar}{\atrm}\btrm}
  }
  \qquad
  \linfer{}{
    \okamperr{}{\atrm}
  }{
    \okamperr{}{\pchk{\atrm}}
  }
  \qquad
  \linfer{}{
    \okamperr{}{\atrm}
  }{
    \okamperr{}{\pmod{\arm}{\atrm}}
  }
\end{align*}

Using this augmented language, unjustified rights amplification is
noted as an error.  To prevent such errors, we modify the typing
system to have judgments of the form
$\soktrm[\crol]{\aenv}{\atrm}{\atyp}$, where $\crol$ indicates the
accumulated guards on a term which must be discharged before the term
may be executed; since $\atrm$ is guarded by $\crol$, it may include
subterms of the form $\pmodup{\arol}{\cdot}$ when
$\crol\rsup\ramplify{\arol}$.  In addition to adding rules for checked
role modifiers, we also modify \RN{t-grd} and \RN{t-mod-up}. The rule
\RN{t-mod-up} ensures that any amplification is justified by $\crol$.
The rule \RN{t-grd} allows guards to be used in checking guarded
terms; the rule is sound since guarded terms must be checked before
they are executed.
\begin{align*}&
  \linfer{t-grd\PRIME}{
    \soktrm[\crol\rmore\arol]{\aenv}{\atrm}{\atyp}
  }{
    \soktrm[\crol]{\aenv}{\pgrd{\arol}{\atrm}}{\tgrd{\arol}{\atyp}}
  }
  &&
  \linferSIDE{t-mod-up\PRIME}{
    \soktrm[\crol]{\aenv}{\atrm}{\tblk{\brol}{\atyp}}
  }{
    \soktrm[\crol]{\aenv}{\pmodup{\arol}{\atrm}}{\tblk{\brol\rminus\arol}{\atyp}}
  }{
    \crol\rsup\ramplify{\arol}
  }
  \\&
  \linferSIDE{t-mod-dn-checked}{
    \soktrm[\crol]{\aenv}{\atrm}{\tblk{\brol}{\atyp}}
  }{
    \soktrm[\crol]{\aenv}{\pmoddn[\drol]{\arol}{\atrm}}{\tblk{\brol}{\atyp}}
  }{
    \text{if}\; \theab=\thea 
    \\ 
    \text{then}\; \okrsup{}{\arol}{\brol}
  }
  &&
  \linferSIDE{t-mod-up-checked}{
    \soktrm[\crol]{\aenv}{\atrm}{\tblk{\brol}{\atyp}}
  }{
    \soktrm[\crol]{\aenv}{\pmodup[\drol]{\arol}{\atrm}}{\tblk{\brol\rminus\arol}{\atyp}}
  }{
    \crol\rmore\drol\rsup\ramplify{\arol}
  }
\end{align*}
One may not assume that top level terms have been guarded; therefore,
let $\soktrm{\aenv}{\atrm}{\atyp}$ be shorthand for
$\soktrm[\rleast]{\aenv}{\atrm}{\atyp}$.

\begin{example}
  \label{selinuxagain}
  \begin{sfmath}
    The functions $\cpsedomtrans{}{}{}$ and $\cpseassign{}$ from
    \autoref{ex:selinux} are not typable using this more restrictive
    system.  Recall the definitions:
    \begin{align*}
      \cpsedomtrans{\arol}{\erol}{\brol} &\eqdef
      \pabsabs{\afun}{\avar}{
        \pchk{\pgrd{\arol}{\punit}} 
        \SEMI
        \pappapp{\afun}{
          \pgrd{\erol}{\pabsabs{\bfun}{\bvar}{\pmodeq{\brol}{\papp{\bfun}{\bvar}}}}}{
          \avar}
      }
      \\
      \cpseassign{\erol} &\eqdef
      \pabsabs{\afun}{\avar}{
        \pabs{\bvar}{
          \plet{\bfun}{\pmodeq{\erol}{\pchk{\avar}}}
          \pappapp{\bfun}{\afun}{\bvar}
        }
      }
    \end{align*}
    The amplification of $\brol$ in $\cpsedomtrans{}{}{}$ is not
    justified; neither is the amplification of $\erol$ in
    $\cpseassign{}$.  The required form is:
    \begin{align*}
      \cpsedomtrans{\arol}{\erol}{\brol} &\eqdef
      \pgrd{\ramplify{\brol}}{\pabsabs{\afun}{\avar}{
        \pchk{\pgrd{\arol}{\punit}} 
        \SEMI
        \pappapp{\afun}{
          \pgrd{\erol}{\pabsabs{\bfun}{\bvar}{\pmodeq{\brol}{\papp{\bfun}{\bvar}}}}}{
          \avar}
      }}
      \\
      \cpseassign{\erol} &\eqdef
      \pgrd{\ramplify{\erol}}{\pabsabs{\afun}{\avar}{
        \pabs{\bvar}{
          \plet{\bfun}{\pmodeq{\erol}{\pchk{\avar}}}
          \pappapp{\bfun}{\afun}{\bvar}
        }
      }}
    \end{align*}
    The login example must now be modified in order to discharge the
    guards.  Again the modifications are straightforward:
    \begin{displaymath}
      \begin{array}{l}
        \plet{\setranslu}{\pchk{\cpsedomtrans{\rlogin}{\xuser}{\ruser}}}{} \\
        \plet{\setransdl}{\pchk{\cpsedomtrans{\rdaemon}{\xlogin}{\rlogin}}}{} \\
        \plet{\seassignu}{\pchk{\cpseassign{\xuser}}}{} \\
        \plet{\seassignl}{\pchk{\cpseassign{\xlogin}}}{} \\
        \plet{\seshell}{\papp{\seassignu}{\pabsp{}{\atrm}}} \\
        \pletNOSEMI{\selogin}{
          \papp{\seassignl}{
            (\pabs{\sepwd}{
              \pcond{\sepwd\eqstr\text{"secret"}}{\papp{\papp{\setranslu}{\seshell}}{\punit}}{\ldots}
              )\SEMI
            }}}
        \\
        \pmodeq{\rdaemon}{\btrm}
      \end{array}
    \end{displaymath}
    Thus modified, the program types correctly, but will only execute
    in a context that dominates the four roles $\ramplify{\ruser}$,
    $\ramplify{\xuser}$, $\ramplify{\rlogin}$, and
    $\ramplify{\xlogin}$.  This ensures that domain transitions and
    assignments are created by authorized code. \qed
  \end{sfmath}
\end{example}

\autoref{amperr} establishes that the typing system is sufficient to
prevent role modification errors.  The proof of
\autoref{amperr} relies on the following lemma, which establishes the
relation between typing and $\fmark{}{}$.
\begin{lemma}
  If $\soktrm[\crol\rmore\arol]{\aenv}{\atrm}{\atyp}$
  then $\soktrm[\crol]{\aenv}{\fmark{\arol}{\atrm}}{\atyp}$.
  \begin{proof}
    By induction on the derivation of the typing judgment, appealing to
    the definition of $\fmark{}{}$.
  \end{proof}
\end{lemma}
\begin{proposition}
  \label{amperr}
  If $\soktrm{}{\atrm}{\atyp}$ and $\okevals{\arol}{\atrm}{\btrm}$ then
  $\lnot(\okamperr{}{\btrm})$
  \begin{proofsketch}
    That $\lnot(\okamperr{}{\atrm})$ follows immediately from the
    definition of role modification error, combined with
    \RN{t-mod-up\PRIME} and \RN{t-mod-up-checked}.  It remains only to
    show that typing is preserved by reduction.  We prove this for the
    type systems of \autoref{sec:typing} in the next section.  The
    proof extends easily to the type system considered here.  The only
    wrinkle is the evaluation rule for $\pchk{}{}$, which is handled
    using the previous lemma.
  \end{proofsketch}
\end{proposition}

%

%
%

%
\section{Proof of Type Safety Theorems}
\label{sec:proofs}

The proofs for the first and second systems are similar, both relying
on well-studied techniques \cite{TAPL}.  We present proofs for the
second system, which is the more challenging of the two.

\begin{definition}[Compatibility]
  Types $\atyp$ and $\btyp$ are \emph{compatible} (notation $\compatible{\atyp}{\btyp}$) if
  $\atyp=\btyp$ or $\atyp=\tblk{\arol}{\ctyp}$
  and $\btyp=\tblk{\brol}{\ctyp}$, for some type $\ctyp$.
  \qed
\end{definition}

The following lemmas have straightforward inductive proofs.
\begin{lemma}[Compatibility]
  \label{result:commontype}
  If $\soktsub{}{\atyp}{\atyp'}$ then $\compatible{\atyp}{\btyp}$
  iff $\compatible{\atyp'}{\btyp}$.
  \qed
\end{lemma}

\begin{lemma}[Substitution]
  \label{result:subst}
  If $\soktrm{\aenv}{\atrm}{\atyp}$ and
  $\soktrm{\aenv, \avar:\atyp}{\btrm}{\btyp}$, then
  $\soktrm{\aenv}{\subst{\btrm}{\avar}{\atrm}}{\btyp}$.
  \qed
\end{lemma}

\begin{lemma}[Bound Weakening]
  \label{result:envsub}
  If $\soktrm{\aenv,\VAR{\avar}{\btyp}}{\atrm}{\atyp}$ and
  $\soktsub{}{\btyp'}{\btyp}$, then
  $\soktrm{\aenv, \avar:\btyp'}{\atrm}{\atyp}$.
  \qed
\end{lemma}


\begin{lemma}[Canonical Forms]
  \label{result:second:canonical}
  \leavevmode
  \begin{enumerate}[\em(1)]
  \item\label{item:second:canonical:func} If
    $\boktrm{}{\aval}{\tabs{\atyp}{\btyp}}$ then $\aval$ has form
    $\pabsp{\avar}{\atrm}$ where $\boktrm{\VAR\avar\atyp}{\atrm}{\btyp}$.
  \item\label{item:second:canonical:comp} If
    $\boktrm{}{\aval}{\tblk{\arol}{\atyp}}$ then $\aval$ has form
    $\pblk{\atrm}$ where $\boktrm{}{\atrm}{\atyp}$ and
    $\arol=\rleast$.
  \item\label{item:second:canonical:guard} If
    $\boktrm{}{\aval}{\tgrd{\arol}{\atyp}}$ then $\aval$ has form
    $\pgrd{\brol}{\atrm}$ where $\boktrm{}{\atrm}{\atyp}$ and
    $\okrsup{}{\brol}{\arol}$.
  \end{enumerate}
\end{lemma}
\begin{proof}
  \leavevmode
  \begin{enumerate}[(1)]
  \item
    By induction on derivation
    of $\boktrm{}{\aval}{\tabs{\atyp}{\btyp}}$.  The only applicable
    cases are \RN{t-sub} and \RN{t-abs}.

    \begin{proofcases}
      \proofcase{\RN{t-sub}} We know
      $\boktrm{}{\aval}{\tabs{\atyp'}{\btyp'}}$, where
      $\boktrm{}{\aval}{\tabs{\atyp}{\btyp}}$ and
      $\boktsub{}{\tabs{\atyp}{\btyp}}{\tabs{\atyp'}{\btyp'}}$, so
      $\boktsub{}{\atyp'}{\atyp}$ and $\boktsub{}{\btyp}{\btyp'}$.  By the
      IH, $\aval$ has form $\pabsp{\avar}{\atrm}$ where
      $\boktrm{\VAR\avar\atyp}{\atrm}{\btyp}$.  By
      \autoref{result:envsub} and subsumption,
      $\boktrm{\VAR\avar\atyp'}{\atrm}{\btyp'}$.

      \proofcase{\RN{t-abs}} Immediate.
    \end{proofcases}
  \item
    By induction on derivation of
    $\boktrm{}{\aval}{\tblk{\arol}{\atyp}}$.  The only applicable cases
    are \RN{t-sub} and \RN{t-unit}.
    \begin{proofcases}
      \proofcase{\RN{t-sub}} We know
      $\boktrm{}{\aval}{\tblk{\arol'}{\atyp'}}$, where
      $\boktrm{}{\aval}{\tblk{\arol}{\atyp}}$ and
      $\boktsub{}{\tblk{\arol}{\atyp}}{\tblk{\arol'}{\atyp'}}$, so
      $\boktsub{}{\atyp}{\atyp'}$ and $\okrsup{}{\arol}{\arol'}$.  By the
      IH, $\aval$ has form $\pblk{\atrm}$ where $\boktrm{}{\atrm}{\atyp}$
      and $\arol=\rleast$, so $\arol'=\rleast$.  By subsumption,
      $\boktrm{}{\atrm}{\atyp'}$.

      \proofcase{\RN{t-unit}} Immediate.
    \end{proofcases}
  \item
    By induction on derivation of
    $\boktrm{}{\aval}{\tgrd{\arol}{\atyp}}$.  The only applicable
    cases are \RN{t-sub} and \RN{t-grd}.
    \begin{proofcases}

      \proofcase{\RN{t-sub}} We know
      $\boktrm{}{\aval}{\tgrd{\arol'}{\atyp'}}$, where
      $\boktrm{}{\aval}{\tgrd{\arol}{\atyp}}$ and
      $\boktsub{}{\tgrd{\arol}{\atyp}}{\tgrd{\arol'}{\atyp'}}$, so
      $\boktsub{}{\atyp}{\atyp'}$ and $\okrsup{}{\arol}{\arol'}$.  By the
      IH, $\aval$ has form $\pgrd{\brol}{\atrm}$ where
      $\boktrm{}{\atrm}{\atyp}$ and $\okrsup{}{\brol}{\arol}$, so
      $\okrsup{}{\brol}{\arol'}$.  By subsumption,
      $\boktrm{}{\atrm}{\atyp'}$.

      \proofcase{\RN{t-grd}} Immediate.
      \popQED
    \end{proofcases}
  \end{enumerate}
\end{proof}

\begin{proposition}[Preservation]
  \label{result:second:preservation}
  If $\boktrm{}{\atrm}{\atyp}$ and $\okeval{\arol}{\atrm}{\btrm}$ then
  there exists $\btyp$ such that $\compatible{\btyp}{\atyp}$ and
  $\boktrm{}{\btrm}{\btyp}$ and if $\dominates{\arol}{\btyp}$ then
  $\dominates{\arol}{\atyp}$.
\end{proposition}


\begin{proof}
  By induction on the derivation of $\boktrm{}{\atrm}{\atyp}$.  The
  induction hypothesis includes the quantification over $\arol$,
  $\btrm$.  For values, the result is trivial; thus we consider only
  the rules for non-values.
  \begin{proofcases}

    \proofcase{\RN{t-sub}} We know $\boktrm{}{\atrm}{\atyp'}$, where
    $\boktrm{}{\atrm}{\atyp}$ and $\boktsub{}{\atyp}{\atyp'}$, and
    $\okeval{\arol}{\atrm}{\btrm}$.  Applying the IH to
    $\boktrm{}{\atrm}{\atyp}$ and $\okeval{\arol}{\atrm}{\btrm}$ yields
    $\btyp$ such that $\boktrm{}{\btrm}{\btyp}$ and
    $\compatible{\btyp}{\atyp}$ and if $\dominates{\arol}{\btyp}$ then
    $\dominates{\arol}{\atyp}$.  By \autoref{result:dominates}\ref{result:second:subdominates},
    this extends to if $\dominates{\arol}{\btyp}$ then
    $\dominates{\arol}{\atyp'}$.  In addition, by
    \autoref{result:commontype}, we have $\compatible{\btyp}{\atyp'}$.


    \proofcase{\RN{t-app}} We know
    $\boktrm{}{\papp{\atrm}{\btrm}}{\atyp_2}$, where
    $\boktrm{}{\atrm}{\tabs{\atyp_1}{\atyp_2}}$ and
    $\boktrm{}{\btrm}{\atyp_1}$, and
    $\okeval{\arol}{\papp{\atrm}{\btrm}}{\ctrm}$.  There are two
    subcases depending on the reduction rule used in
    $\okeval{\arol}{\papp{\atrm}{\btrm}}{\ctrm}$.

    \begin{proofsubcases}
      \proofsubcase{$\atrm$ is a value} By
      \autoref{result:second:canonical}, $\atrm =
      \pabs{\avar}{\atrm'}$ and
      $\boktrm{\VAR{\avar}{\atyp_1}}{\atrm'}{\atyp_2}$.  The reduction yields
      $\ctrm = \subst{\atrm'}{\avar}{\btrm}$.  By
      \autoref{result:subst}, $\boktrm{}{\ctrm}{\atyp_2}$.  The
      remaining requirements on $\atyp_2$ are immediate.

      \proofsubcase{$\atrm$ has a reduction} Therefore
      $\okeval{\arol}{\atrm}{\atrm'}$ and $\ctrm =
      \papp{\atrm'}{\btrm}$.  Applying the IH to
      $\boktrm{}{\atrm}{\tabs{\atyp_1}{\atyp_2}}$ and
      $\okeval{\arol}{\atrm}{\atrm'}$ yields $\btyp$ such that
      $\boktrm{}{\atrm'}{\btyp}$ and
      $\compatible{\btyp}{\tabs{\atyp_1}{\atyp_2}}$, which implies
      that $\btyp = \tabs{\atyp_1}{\atyp_2}$.  Hence
      $\boktrm{}{\ctrm}{\atyp_2}$.  The remaining requirements on
      $\atyp_2$ are immediate.
    \end{proofsubcases}


    \proofcase{\RN{t-fix}} We know
    $\boktrm{}{\pfix{\atrm}}{\atyp}$, where
    $\boktrm{}{\atrm}{\tabs{\atyp}{\atyp}}$ and
    $\okeval{\arol}{\pfix{\atrm}}{\ctrm}$.  There are two
    subcases depending on the reduction rule used in
    $\okeval{\arol}{\pfix{\atrm}}{\ctrm}$.

    \begin{proofsubcases}
      \proofsubcase{$\atrm$ is a value} By
      \autoref{result:second:canonical}, $\atrm =
      \pabs{\avar}{\atrm'}$ and
      $\boktrm{\VAR{\avar}{\atyp}}{\atrm'}{\atyp}$.  The reduction yields
      $\ctrm = \subst{\atrm'}{\avar}{\atrm}$.  By
      \autoref{result:subst}, $\boktrm{}{\ctrm}{\atyp}$.  The
      remaining requirements on $\atyp$ are immediate.
      
      \proofsubcase{$\atrm$ has a reduction} Therefore
      $\okeval{\arol}{\atrm}{\atrm'}$ and $\ctrm =
      \pfix{\atrm'}$.  Applying the IH to
      $\boktrm{}{\atrm}{\tabs{\atyp}{\atyp}}$ and
      $\okeval{\arol}{\atrm}{\atrm'}$ yields $\btyp$ such that
      $\boktrm{}{\atrm'}{\btyp}$ and
      $\compatible{\btyp}{\tabs{\atyp}{\atyp}}$, which implies
      that $\btyp = \tabs{\atyp}{\atyp}$.  Hence
      $\boktrm{}{\ctrm}{\atyp}$.  The remaining requirements on
      $\atyp$ are immediate.
    \end{proofsubcases}


    \proofcase{\RN{t-chk}} We know
    $\boktrm{}{\pchk{\atrm}}{\tblk{\arol_1}{\atyp}}$, where
    $\boktrm{}{\atrm}{\tgrd{\arol_1}{\atyp}}$, and
    $\okeval{\arol}{\pchk{\atrm}}{\ctrm}$.  There are two subcases
    depending on the reduction rule used in
    $\okeval{\arol}{\pchk{\atrm}}{\ctrm}$.

    \begin{proofsubcases}
      \proofsubcase{$\atrm$ is a value} By
      \autoref{result:second:canonical}, $\atrm =
      \pgrd{\arol_2}{\atrm'}$ and $\boktrm{}{\atrm'}{\atyp}$ and
      $\okrsub{}{\arol_1}{\arol_2}$.  The reduction yields $\ctrm =
      \pblk{\atrm'}$ and from the reduction we deduce
      $\okrsup{}{\arol}{\arol_2}$, so $\okrsup{}{\arol}{\arol_1}$
      always holds.  We assign type
      $\boktrm{}{\ctrm}{\tblk{\rleast}{\atyp}}$, where
      $\compatible{\tblk{\rleast}{\atyp}}{\tblk{\arol_1}{\atyp}}$, and
      we have already shown that $\okrsup{}{\arol}{\rleast}$ implies
      $\okrsup{}{\arol}{\arol_1}$.

      \proofsubcase{$\atrm$ has a reduction} Therefore
      $\okeval{\arol}{\atrm}{\atrm'}$ and $\ctrm = \pchk{\atrm'}$.
      Applying the IH to $\boktrm{}{\atrm}{\tgrd{\arol_1}{\atyp}}$ and
      $\okeval{\arol}{\atrm}{\atrm'}$ yields $\btyp$ such that
      $\boktrm{}{\atrm'}{\btyp}$ and
      $\compatible{\btyp}{\tgrd{\arol_1}{\atyp}}$, so $\btyp =
      \tgrd{\arol_1}{\atyp}$.  Hence
      $\boktrm{}{\ctrm}{\tblk{\arol_1}{\atyp}}$.  The remaining
      requirements on $\tblk{\arol_1}{\atyp}$ are immediate.
    \end{proofsubcases}


    \proofcase{\RN{t-bind}}
    We know
    $\boktrm{}{\plet{\avar}{\atrm}\btrm}{\tblk{\arol_1\rmore\arol_2}{\atyp_2}}$,
    where $\boktrm{}{\atrm}{\tblk{\arol_1}{\atyp_1}}$ and
    $\boktrm{\VAR{\avar}{\atyp_1}}{\btrm}{\tblk{\arol_2}{\atyp_2}}$, and
    $\okeval{\arol}{\plet{\avar}{\atrm}\btrm}{\ctrm}$.  There are two
    subcases depending on the reduction rule used in
    $\okeval{\arol}{\plet{\avar}{\atrm}\btrm}{\ctrm}$.

    \begin{proofsubcases}
      \proofsubcase{$\atrm$ is a value} By
      \autoref{result:second:canonical}, $\atrm = \pblk{\atrm'}$ and
      $\boktrm{}{\atrm'}{\atyp_1}$ and $\arol_1 = \rleast$, so
      $\arol_1\rmore\arol_2=\arol_2$.  The reduction yields $\ctrm =
      \subst{\btrm}{\avar}{\atrm'}$.  By \autoref{result:subst},
      $\boktrm{}{\ctrm}{\tblk{\arol_2}{\atyp_2}}$.  The remaining
      requirements on $\atyp_2$ are immediate.

      \proofsubcase{$\atrm$ has a reduction} Therefore
      $\okeval{\arol}{\atrm}{\atrm'}$ and $\ctrm =
      \plet{\avar}{\atrm'}\btrm$.  Applying the IH to
      $\boktrm{}{\atrm}{\tblk{\arol_1}{\atyp_1}}$ and
      $\okeval{\arol}{\atrm}{\atrm'}$ yields $\btyp$ such that
      $\boktrm{}{\atrm'}{\btyp}$ and
      $\compatible{\btyp}{\tblk{\arol_1}{\atyp_1}}$ and if
      $\dominates{\arol}{\btyp}$ then
      $\dominates{\arol}{\tblk{\arol_1}{\atyp_1}}$.  Hence $\btyp =
      \tblk{\arol_3}{\atyp_1}$, for some $\arol_3$, and
      $\okrsup{}{\arol}{\arol_3}$ implies $\okrsup{}{\arol}{\arol_1}$.
      We deduce
      $\boktrm{}{\ctrm}{\tblk{\arol_3\rmore\arol_2}{\atyp_2}}$, where
      $\compatible{\tblk{\arol_3\rmore\arol_2}{\atyp_2}}{\tblk{\arol_1\rmore\arol_2}{\atyp_2}}$.
      Finally, suppose
      $\dominates{\arol}{\tblk{\arol_3\rmore\arol_2}{\atyp_2}}$, i.e.,
      $\okrsup{}{\arol}{\arol_3\rmore\arol_2}$, so
      $\okrsup{}{\arol}{\arol_3}$ and $\okrsup{}{\arol}{\arol_2}$.  By
      the above, this entails $\okrsup{}{\arol}{\arol_1}$, so
      $\okrsup{}{\arol}{\arol_1\rmore\arol_2}$.  Therefore
      $\dominates{\arol}{\tblk{\arol_1\rmore\arol_2}{\atyp_2}}$, as
      required.
    \end{proofsubcases}


    \proofcase{\RN{t-mod-up}} We know
    $\boktrm{}{\pmodup{\arol_1}{\atrm}}{\tblk{\arol_2\rminus{\arol_1}}{\atyp}}$,
    where $\boktrm{}{\atrm}{\tblk{\arol_2}{\atyp}}$, and
    $\okeval{\arol}{\pmodup{\arol_1}{\atrm}}{\ctrm}$.  There are two
    subcases depending on the reduction rule used in
    $\okeval{\arol}{\pmodup{\arol_1}{\atrm}}{\ctrm}$.

    \begin{proofsubcases}
      \proofsubcase{$\atrm$ is a value} Therefore $\ctrm = \atrm$ and
      $\boktrm{}{\ctrm}{\tblk{\arol_2}{\atyp}}$, where
      $\compatible{\tblk{\arol_2}{\atyp}}{\tblk{\arol_2\rminus{\arol_1}}{\atyp}}$.
      By \autoref{result:second:canonical}, $\atrm = \pblk{\atrm'}$
      and $\boktrm{}{\atrm'}{\atyp}$ and $\arol_2 = \rleast$, and the
      remaining requirement on $\tblk{\arol_2}{\atyp}$, that
      $\dominates{\arol}{\tblk{\arol_2}{\atyp}}$ implies
      $\dominates{\arol}{\tblk{\arol_2\rminus{\arol_1}}{\atyp}}$, is
      immediate.

      \proofsubcase{$\atrm$ has a reduction} Therefore
      $\okeval{\rmup{\arol_1}{\arol}}{\atrm}{\atrm'}$ and $\ctrm =
      \pmodup{\arol_1}{\atrm'}$.  Applying the IH to
      $\boktrm{}{\atrm}{\tblk{\arol_2}{\atyp}}$ and
      $\okeval{\rmup{\arol_1}{\arol}}{\atrm}{\atrm'}$ yields $\btyp$
      such that $\boktrm{}{\atrm'}{\btyp}$ and
      $\compatible{\btyp}{\tblk{\arol_2}{\atyp}}$, so $\btyp =
      \tblk{\arol_3}{\atyp}$ for some $\arol_3$, and if
      $\dominates{\rmup{\arol_1}{\arol}}{\btyp}$ then
      $\dominates{\rmup{\arol_1}{\arol}}{\tblk{\arol_2}{\atyp}}$,
      i.e., $\okrsup{}{\arol \rmore \arol_1}{\arol_3}$ implies
      $\okrsup{}{\arol \rmore \arol_1}{\arol_2}$.  We have
      $\boktrm{}{\pmodup{\arol_1}{\atrm'}}{\tblk{\arol_3\rminus{\arol_1}}{\atyp}}$
      and
      $\compatible{\tblk{\arol_3\rminus{\arol_1}}{\atyp}}{\tblk{\arol_2\rminus{\arol_1}}{\atyp}}$.
      Finally, if
      $\dominates{\arol}{\tblk{\arol_3\rminus{\arol_1}}{\atyp}}$, then
      $\okrsup{}{\arol}{\arol_3\rminus{\arol_1}}$, so
      $\okrsup{}{\arol\rmore\arol_1}{(\arol_3\rminus{\arol_1})\rmore\arol_1}
      = (\arol_3\rmore\arol_1)\rless\rmost = \arol_3\rmore\arol_1$.
      Hence $\okrsup{}{\arol\rmore\arol_1}{\arol_3}$, so
      $\okrsup{}{\arol\rmore\arol_1}{\arol_2}$, and
      $\arol\rminus{\arol_1} = (\arol\rminus{\arol_1})\rmore\rleast =
      \okrsup{}{(\arol\rmore\arol_1)\rminus{\arol_1}}{\arol_2\rminus{\arol_1}}$.
      Therefore $\okrsup{}{\arol}{\arol_2\rminus{\arol_1}}$ and
      $\dominates{\arol}{\tblk{\arol_2\rminus{\arol_1}}{\atyp}}$, as
      required.
    \end{proofsubcases}


    \proofcase{\RN{t-mod-dn}}
    We know $\boktrm{}{\pmoddn{\arol_1}{\atrm}}{\tblk{\arol_2}{\atyp}}$,
    where $\boktrm{}{\atrm}{\tblk{\arol_2}{\atyp}}$, and
    $\okeval{\arol}{\pmoddn{\arol_1}{\atrm}}{\ctrm}$.  There are two
    subcases depending on the reduction rule used in
    $\okeval{\arol}{\pmoddn{\arol_1}{\atrm}}{\ctrm}$.

    \begin{proofsubcases}
      \proofsubcase{$\atrm$ is a value} Therefore $\ctrm = \atrm$ and
      $\boktrm{}{\ctrm}{\tblk{\arol_2}{\atyp}}$, and we are done.

      \proofsubcase{$\atrm$ has a reduction} Therefore
      $\okeval{\rmdn{\arol_1}{\arol}}{\atrm}{\atrm'}$ and $\ctrm =
      \pmoddn{\arol_1}{\atrm'}$.  By
      $\okrsup{}{\arol}{\rmdn{\arol_1}{\arol}}$ and
      \autoref{result:rednhigher}, we have
      $\okeval{\arol}{\atrm}{\atrm'}$.  Applying the IH to
      $\boktrm{}{\atrm}{\tblk{\arol_2}{\atyp}}$ and
      $\okeval{\arol}{\atrm}{\atrm'}$ yields $\btyp$ such that
      $\boktrm{}{\atrm'}{\btyp}$ and
      $\compatible{\btyp}{\tblk{\arol_2}{\atyp}}$, so $\btyp =
      \tblk{\arol_3}{\atyp}$ for some $\arol_3$, and if
      $\dominates{\arol}{\btyp}$ then
      $\dominates{\arol}{\tblk{\arol_2}{\atyp}}$.  Hence
      $\boktrm{}{\pmoddn{\arol_1}{\atrm'}}{\tblk{\arol_3}{\atyp}}$,
      which completes the subcase.
      \popQED
    \end{proofsubcases}
  \end{proofcases}
\end{proof}

\begin{corollary}
  \label{result:second:rednvalue}
  If $\boktrm{}{\atrm}{\atyp}$ and $\okevals{\arol}{\atrm}{\aval}$,
  then $\dominates{\arol}{\atyp}$.
\end{corollary}
\begin{proof}
  By induction on the length of the reduction sequence
  $\okevals{\arol}{\atrm}{\aval}$.  For the base case, $\atrm=\aval$
  and \autoref{result:second:canonical} implies that
  $\dominates{\arol}{\atyp}$, because every non-computation type is
  dominated by any role, and in a computation type $\atyp =
  \tblk{\brol}{\btyp}$ \autoref{result:second:canonical} tells us that
  $\brol=\rleast$.  For the inductive step, there exists $\btrm$ such
  that $\okeval{\arol}{\atrm}{\btrm}$ and
  $\okevals{\arol}{\btrm}{\aval}$.  By
  \autoref{result:second:preservation}, there exists $\btyp$ such that
  $\boktrm{}{\btrm}{\btyp}$ and if $\dominates{\arol}{\btyp}$ then
  $\dominates{\arol}{\atyp}$.  Applying the IH to
  $\boktrm{}{\btrm}{\btyp}$ and $\okevals{\arol}{\btrm}{\aval}$ yields
  $\dominates{\arol}{\btyp}$, hence $\dominates{\arol}{\atyp}$ as
  required.
\end{proof}

\begin{proposition}[Progress]
  \label{result:second:progress}
  For all $\arol$, 
  if $\boktrm{}{\atrm}{\atyp}$ then either $\atrm$ is a value,
  $\okerr{\arol}{\atrm}$, or there exists $\btrm$ such that
  $\okeval{\arol}{\atrm}{\btrm}$.
\end{proposition}
\begin{proof}
  By induction on the derivation of $\boktrm{}{\atrm}{\atyp}$.  We
  need only consider the cases when $\atrm$ is not a value.
  \begin{proofcases}

    \proofcase{\RN{t-sub}} We know $\boktrm{}{\atrm}{\atyp'}$, where
    $\boktrm{}{\atrm}{\atyp}$ and $\boktsub{}{\atyp}{\atyp'}$.
    Immediate by the IH.


    \proofcase{\RN{t-app}} We know
    $\boktrm{}{\papp{\atrm}{\btrm}}{\atyp_2}$, where
    $\boktrm{}{\atrm}{\tabs{\atyp_1}{\atyp_2}}$ and
    $\boktrm{}{\btrm}{\atyp_1}$.  Apply the IH to
    $\boktrm{}{\atrm}{\tabs{\atyp_1}{\atyp_2}}$ and role $\arol$.  If
    $\atrm$ is a value, then, by \autoref{result:second:canonical},
    $\atrm$ has form $\pabsp{\avar}{\ctrm}$, so
    $\okeval{\arol}{\papp{\atrm}{\btrm}}{\subst{\ctrm}{\avar}{\btrm}}$.
    If $\okerr{\arol}{\atrm}$, then
    $\okerr{\arol}{\papp{\atrm}{\btrm}}$.  Finally, if
    $\okeval{\arol}{\atrm}{\ctrm}$, then
    $\okeval{\arol}{\papp{\atrm}{\btrm}}{\papp{\ctrm}{\btrm}}$.


    \proofcase{\RN{t-fix}} We know
    $\boktrm{}{\pfix{\atrm}}{\atyp}$, where
    $\boktrm{}{\atrm}{\tabs{\atyp}{\atyp}}$.  Apply the IH to
    $\boktrm{}{\atrm}{\tabs{\atyp}{\atyp}}$ and role $\arol$.  If
    $\atrm$ is a value, then, by \autoref{result:second:canonical},
    $\atrm$ has form $\pabsp{\avar}{\ctrm}$, so
    $\okeval{\arol}{\pfix{\atrm}}{\subst{\ctrm}{\avar}{\pabsp{\avar}{\ctrm}}}$.
    If $\okerr{\arol}{\atrm}$, then
    $\okerr{\arol}{\pfix{\atrm}}$.  Finally, if
    $\okeval{\arol}{\atrm}{\ctrm}$, then
    $\okeval{\arol}{\pfix{\atrm}}{\pfix{\ctrm}}$.


    \proofcase{\RN{t-chk}} We know
    $\boktrm{}{\pchk{\atrm}}{\tblk{\arol_1}{\atyp}}$, where
    $\boktrm{}{\atrm}{\tgrd{\arol_1}{\atyp}}$.  Apply the IH to
    $\boktrm{}{\atrm}{\tgrd{\arol_1}{\atyp}}$ and role $\arol$.  If
    $\atrm$ is a value, then, by \autoref{result:second:canonical},
    there exists $\brol$, $\ctrm$ such that $\atrm=\pgrd{\brol}{\ctrm}$,
    so either $\okeval{\arol}{\pchk{\atrm}}{\pblk{\ctrm}}$ or
    $\okerr{\arol}{\pchk{\atrm}}$ depending on whether
    $\okrsup{}{\arol}{\brol}$ holds or not.  If $\okerr{\arol}{\atrm}$,
    then $\okerr{\arol}{\pchk{\atrm}}$.  Finally, if
    $\okeval{\arol}{\atrm}{\ctrm}$, then
    $\okeval{\arol}{\pchk{\atrm}}{\pchk{\ctrm}}$.


    \proofcase{\RN{t-bind}} We know
    $\boktrm{}{\plet{\avar}{\atrm}\btrm}{\tblk{\arol_1\rmore\arol_2}{\atyp_2}}$,
    where $\boktrm{}{\atrm}{\tblk{\arol_1}{\atyp_1}}$ and
    $\boktrm{\VAR{\avar}{\atyp_1}}{\btrm}{\tblk{\arol_2}{\atyp_2}}$.
    Apply the IH to $\boktrm{}{\atrm}{\tblk{\arol_1}{\atyp_1}}$ and role
    $\arol$.  If $\atrm$ is a value, then, by
    \autoref{result:second:canonical}, $\atrm$ has form $\pblk{\ctrm}$,
    so
    $\okeval{\arol}{\plet{\avar}{\atrm}\btrm}{\subst{\btrm}{\avar}{\ctrm}}$.
    If $\okerr{\arol}{\atrm}$, then
    $\okerr{\arol}{\plet{\avar}{\atrm}\btrm}$.  Finally, if
    $\okeval{\arol}{\atrm}{\ctrm}$, then
    $\okeval{\arol}{\plet{\avar}{\atrm}\btrm}{\plet{\avar}{\ctrm}\btrm}$.


    \proofcase{\RN{t-mod-up}} We know
    $\boktrm{}{\pmodup{\arol_1}{\atrm}}{\tblk{\arol_2\rminus{\arol_1}}{\atyp}}$,
    where $\boktrm{}{\atrm}{\tblk{\arol_2}{\atyp}}$.  Apply the IH to
    $\boktrm{}{\atrm}{\tblk{\arol_2}{\atyp}}$ and role
    $\rmup{\arol_1}{\arol}$.  If $\atrm$ is a value, then
    $\okeval{\arol}{\pmodup{\arol_1}{\atrm}}{\atrm}$.  If
    $\okerr{\rmup{\arol_1}{\arol}}{\atrm}$, then
    $\okerr{\arol}{\pmodup{\arol_1}{\atrm}}$.  Finally, if
    $\okeval{\rmup{\arol_1}{\arol}}{\atrm}{\ctrm}$, then
    $\okeval{\arol}{\pmodup{\arol_1}{\atrm}}{\pmodup{\arol_1}{\ctrm}}$.


    \proofcase{\RN{t-mod-dn}} We know
    $\boktrm{}{\pmoddn{\arol_1}{\atrm}}{\tblk{\arol_2}{\atyp}}$, where
    $\boktrm{}{\atrm}{\tblk{\arol_2}{\atyp}}$.  Apply the IH to
    $\boktrm{}{\atrm}{\tblk{\arol_2}{\atyp}}$ and role
    $\rmdn{\arol_1}{\arol}$.  If $\atrm$ is a value, then
    $\okeval{\arol}{\pmoddn{\arol_1}{\atrm}}{\atrm}$.  If
    $\okerr{\rmdn{\arol_1}{\arol}}{\atrm}$, then
    $\okerr{\arol}{\pmoddn{\arol_1}{\atrm}}$.  Finally, if
    $\okeval{\rmdn{\arol_1}{\arol}}{\atrm}{\ctrm}$, then
    $\okeval{\arol}{\pmoddn{\arol_1}{\atrm}}{\pmoddn{\arol_1}{\ctrm}}$.
    \popQED
  \end{proofcases}
\end{proof}

\begin{ntheorem}[\ref{result:second:divergeorerror}]
  If $\boktrm{}{\atrm}{\atyp}$ and
  $\notdominates{\arol}{\atyp}$, then either $\okdvg{\arol}{\atrm}$ or
  there exists $\btrm$ such that $\okevals{\arol}{\atrm}{\btrm}$ and
  $\okerr{\arol}{\btrm}$.
\end{ntheorem}
\begin{proof}
  We use a coinductive argument to construct a reduction sequence that
  is either infinite or terminates with a role check failure.  When
  $\boktrm{}{\atrm}{\atyp}$ and $\notdominates{\arol}{\atyp}$, we know
  that $\atrm$ is not a value by \autoref{result:second:canonical}.
  By \autoref{result:second:progress}, either $\okerr{\arol}{\atrm}$
  or there exists $\btrm$ such that $\okeval{\arol}{\atrm}{\btrm}$.
  In the former case, we are done.  In the latter case, using
  \autoref{result:second:preservation}, there exists $\btyp$ such that
  $\boktrm{}{\btrm}{\btyp}$ and if $\dominates{\arol}{\btyp}$ then
  $\dominates{\arol}{\atyp}$.  However, we know that
  $\notdominates{\arol}{\atyp}$, so $\notdominates{\arol}{\btyp}$, as
  required.
\end{proof}

%
%
%
\section{Conclusions}
\label{sec:end}

The focus of this paper is programmatic approaches, such as
\JAAS/\DOTNET, that use \RBAC. From a software engineering approach
to the design of components, \RBAC\ facilitates a separation
of concerns: the design of the system is carried out in terms of
a role hierarchy with an associated assignment of permissions to
roles, whereas the actual assignment of users to roles takes place
at the time of deployment.

We have presented two methods to aid the design and use of
components that include such access control code.  The first ---
admittedly standard --- technique enables users of code to deduce
the role at which code must be run. The main use of this analysis is
to optimize code by enabling the removal of some dynamic checks. The
second --- somewhat more novel --- analysis calculates the role that
is verified on all execution paths.  This analysis is potentially
useful in validating architectural security requirements by enabling
code designers to deduce the protection guarantees of their code.

We have demonstrated the use of these methods by modeling Domain
Type Enforcement, as used in SELinux.  As future work, we will
explore extensions to role polymorphism and recursive roles
following the techniques of~\cite{291678,AmaCar93}.

\section*{Acknowledgments}
The presentation of the paper has greatly improved thanks to the
comments of the referees.

\bibliographystyle{ieee}
\bibliography{bib}

\appendix
\end{document}